\documentclass[a4paper,11pt]{article}

\usepackage{amsfonts}
\usepackage{amsmath}
\usepackage{amssymb}
\usepackage[titletoc,toc,title]{appendix}
\usepackage{authblk}
\usepackage{braket}
\usepackage{cancel}
\usepackage[font={footnotesize,it}]{caption}
\usepackage{cite}
\usepackage{comment}
\usepackage{enumitem}
\usepackage{epsfig}
\usepackage{float}
\usepackage[margin=2.5cm]{geometry}
\usepackage{graphicx}
\usepackage{epstopdf}
\usepackage{hyperref}
\usepackage{cleveref}
\usepackage{physics}
\usepackage{ragged2e}
\usepackage{soul}
\usepackage{subcaption}
\usepackage{xcolor}

\hypersetup{
citecolor=red,
colorlinks=true,
filecolor=red,
linkcolor=blue,
linktocpage=true,
urlcolor=blue
}

{\makeatletter\g@addto@macro\bfseries{\boldmath}\makeatother}
\numberwithin{equation}{section}

\def\ttbar{$\text{T}\bar{\text{T}}$}
\def\cft#1{CFT$_{#1}$}
\def\ads#1{AdS$_{#1}$}

\begin{document}

\title{Odd entanglement entropy in $\text{T}\bar{\text{T}}$ deformed CFT$_2$s and holography}

\author[]{Debarshi Basu\thanks{\noindent E-mail:~ {\texttt{debarshi@iitk.ac.in}}}}
\author[]{Saikat Biswas\thanks{\noindent E-mail:~ {\texttt{saikatb21@iitk.ac.in}}}}
\author[]{Ankur Dey\thanks{\noindent E-mail:~ {\texttt{ankurd21@iitk.ac.in}}}}
\author[]{Boudhayan Paul\thanks{\noindent E-mail:~ {\texttt{paul@iitk.ac.in}}} }
\author[]{Gautam Sengupta\thanks{\noindent E-mail:~ {\texttt{sengupta@iitk.ac.in}}}}

\affil[]{
Department of Physics,\\
Indian Institute of Technology,\\ 
Kanpur 208 016, India
}	

\date{}

\maketitle

\thispagestyle{empty}

\begin{abstract}

\bigskip

\noindent
We construct a replica technique to perturbatively compute the odd entanglement entropy (OEE) for bipartite mixed states in $\text{T}\bar{\text{T}}$ deformed CFT$_2$s. This framework is then utilized to obtain the leading order correction to the OEE for two disjoint intervals, two adjacent intervals, and a single interval in $\text{T}\bar{\text{T}}$ deformed thermal CFT$_2$s in the large central charge limit. The field theory results are subsequently reproduced in the high temperature limit from holographic computations for the entanglement wedge cross sections in the dual bulk finite cut-off BTZ geometries. We further show that for finite size $\text{T}\bar{\text{T}}$ deformed CFT$_2$s at zero temperature the corrections to the OEE are vanishing to the leading order from both field theory and bulk holographic computations.

\end{abstract}

\clearpage
\tableofcontents
\clearpage

\section{Introduction}
\label{sn_intro}

Quantum entanglement has emerged as a prominent area of research to explore a wide range of physical phenomena spanning several disciplines from quantum many body systems in condensed matter physics to issues of quantum gravity and black holes. The entanglement entropy (EE) has played a crucial role in this endeavor as a measure for characterizing the entanglement of bipartite pure quantum states although it fails to effectively capture mixed state entanglement due to spurious correlations. In this context several mixed state entanglement and correlation measures such as the reflected entropy, entanglement of purification, balanced partial entanglement etc. have been proposed in quantum information theory.

Interestingly it was possible to compute several of these measures through certain replica techniques for bipartite states in two dimensional conformal field theories (\cft{2}s). In this connection the Ryu Takayanagi (RT) proposal \cite{Ryu:2006bv,Ryu:2006ef} quantitatively characterized the holographic entanglement entropy (HEE) of a subsystem in \cft{}s dual to bulk \ads{} geometries through the \ads{}/\cft{} correspondence. This was extended by the Hubeny Rangamani Takayanagi (HRT) proposal \cite{Hubeny:2007xt} which provided a covariant generalization of the RT proposal for time dependent states in \cft{}s dual to non static bulk \ads{} geometries. The RT and HRT proposals were later proved in \cite{Fursaev:2006ih,Casini:2011kv,Lewkowycz:2013nqa,Dong:2016hjy}.

Recently another computable measure for mixed state entanglement known as the odd entanglement entropy (OEE) was proposed by Tamaoka in \cite{Tamaoka:2018ned}. The OEE may be broadly understood as the von Neumann entropy of the partially transposed reduced density matrix of a given subsystem \cite{Tamaoka:2018ned}.\footnote{This is a loose interpretation as the partially transposed reduced density matrix does not represent a physical state and may contain negative eigenvalues \cite{Tamaoka:2018ned}.} The author in \cite{Tamaoka:2018ned} utilized a suitable replica technique to compute the OEE for a bipartite mixed state configuration of two disjoint intervals in a \cft{2}. Interestingly in \cite{Tamaoka:2018ned} the author proposed a holographic duality relating the OEE and the EE to the bulk entanglement wedge cross section (EWCS) for a given bipartite state in the \ads{3}/\cft{2} scenario. For recent developments see \cite{BabaeiVelni:2019pkw,Kusuki:2019evw,Angel-Ramelli:2020wfo,Mollabashi:2020ifv,BabaeiVelni:2020wfl,Berthiere:2020ihq,Dong:2021clv,Sahraei:2021wqn,Ghasemi:2021jiy,Basak:2022gcv}.

On a different note it was demonstrated by Zamolodchikov \cite{Zamolodchikov:2004ce} that \cft{2}s which have undergone an irrelevant deformation by the determinant of the stress tensor (known as \ttbar{} deformations) exhibit exactly solvable energy spectrum and partition function. These theories display non local UV structure and have an infinite number of possible RG flows leading to the same fixed point. A holographic dual for such theories was proposed in \cite{McGough:2016lol} to be a bulk \ads{3} geometry with a finite radial cut-off. This proposal could be substantiated through the matching of the two point function, energy spectrum and the partition function between the bulk and the boundary (see \cite{Shyam:2017znq,Kraus:2018xrn,Cottrell:2018skz,Taylor:2018xcy,Hartman:2018tkw,Shyam:2018sro,Caputa:2019pam,Giveon:2017myj,Asrat:2017tzd} for further developments). The authors in \cite{Donnelly:2018bef,Lewkowycz:2019xse,Chen:2018eqk,Banerjee:2019ewu,Jeong:2019ylz,Murdia:2019fax,Park:2018snf,Asrat:2019end,He:2019vzf,Grieninger:2019zts,Khoeini-Moghaddam:2020ymm} computed the HEE for bipartite pure state configurations in various \ttbar{} deformed dual \cft{}s. Subsequently the authors in \cite{Asrat:2020uib} obtained the reflected entropy and its holographic dual, the EWCS, for bipartite mixed states in \ttbar{} deformed dual \cft{2}s. Recently the entanglement negativity for various bipartite mixed states in \ttbar{} deformed thermal \cft{2}s, and the corresponding holographic dual for bulk finite cut-off BTZ black hole geometries were computed in \cite{Basu:2023bov}.

Motivated by the developments described above, in this article we compute the OEE for various bipartite mixed states in \ttbar{} deformed dual \cft{2}s. For this purpose we construct an appropriate replica technique and a conformal perturbation theory along the lines of \cite{Chen:2018eqk,Jeong:2019ylz,Basu:2023bov} to develop a path integral formulation for the OEE in \ttbar{} deformed \cft{2}s with a small deformation parameter. This perturbative construction is then utilized to compute the first order corrections to the OEE for two disjoint intervals, two adjacent intervals, and a single interval in a \ttbar{} deformed thermal \cft{2} with a small deformation parameter in the large central charge limit. Subsequently we explicitly compute the bulk EWCS for the above mixed state configurations in the \ttbar{} deformed thermal dual \cft{2}s by employing a construction involving embedding coordinates as described in \cite{Kusuki:2019evw}. Utilizing the EWCS obtained we demonstrate that the first order correction to field theory replica technique results for the OEE in the large central charge and the high temperature limit match exactly with the first order correction to the sum of the EWCS and the HEE verifying the holographic duality between the above quantities in the context of \ttbar{} deformed thermal \cft{2}s. Following this we extend our perturbative construction to \ttbar{} deformed finite size \cft{2}s at zero temperature and demonstrate that the leading order corrections to the OEE are vanishing, which is substantiated through bulk holographic computations involving the EWCS.

This article is organized as follows. In \cref{sn_review} we briefly review the basic features of \ttbar{} deformed \cft{2}s and the OEE. In \cref{sn_oee_tt} we develop a perturbative expansion for the OEE in a \ttbar{} deformed \cft{2}. In \cref{sn_ft} this perturbative construction is then employed to obtain the leading order corrections to the OEE for various bipartite states in a \ttbar{} deformed thermal \cft{2}. Following this we explicitly demonstrate the holographic duality for first order corrections between the OEE and the sum of the bulk EWCS and the HEE for these mixed states. Subsequently in \cref{sn_fs} we extend our perturbative analysis to a \ttbar{} deformed finite size \cft{2} at zero temperature and show that the leading order corrections to the OEE are zero. This is later verified through bulk holographic computations. Finally, we summarize our results in \cref{sn_sum} and present our conclusions. Some of the lengthy technical details of our computations have been described in \cref{app_int_ft}.

\section{Review of earlier literature}
\label{sn_review}

\subsection{\ttbar{} deformation in a \cft{2}}
\label{sn_tt}

We begin with a brief review of a two dimensional conformal field theory deformed by the \ttbar{} operator defined as follows \cite{Zamolodchikov:2004ce}
\begin{align}
\left(T\bar{T}\right)=\frac{1}{8}\left(T_{ab}\,T^{ab}-\left(T^a_a\right)^2\right).
\end{align}
It is a double trace composite operator which satisfies the factorization property \cite{Zamolodchikov:2004ce}. The corresponding deformation generates a one parameter family of theories described by a deformation parameter $\mu\,(\geq 0)$ as given by the following flow equation \cite{Zamolodchikov:2004ce,Chen:2018eqk,Jeong:2019ylz}
\begin{align}
\frac{d\mathcal{I}_{\text{QFT}}^{(\mu)}}{d\mu}=\int d^2x\:(T\bar{T})_{\mu}~~,~~\mathcal{I}_{\text{QFT}}^{(\mu)}\Bigg|_{\mu=0}=\mathcal{I}_{\text{CFT}}\,,
\end{align}
where $\mathcal{I}_{\text{QFT}}^{(\mu)}$ and $\mathcal{I}_{\text{CFT}}$ represent the actions of the deformed and undeformed theories respectively. The deformation parameter $\mu$ has dimensions of length squared. Note that the energy spectrum may be determined exactly for a \ttbar{} deformed \cft{2} \cite{Smirnov:2016lqw,Cavaglia:2016oda}.

When $\mu$ is small, the action of the deformed \cft{2} may be perturbatively expanded as \cite{Chen:2018eqk,Jeong:2019ylz}
\begin{align}
\label{def_act}
\mathcal{I}_{\text{QFT}}^{(\mu)}=\mathcal{I}_{\text{CFT}}+\mu \int d^{2}x\;(T\bar{T})_{\mu=0}
=\mathcal{I}_{\text{CFT}}+\mu\int d^{2}x\;\left(T\bar{T}-\Theta^2\right)\,,
\end{align}
where $T\equiv T_{ww}$, $\bar{T}\equiv T_{\bar{w}\bar{w}}$ and $\Theta\equiv T_{w\bar{w}}$ describe the components of the stress tensor of the undeformed theory expressed in the complex coordinates $(w,\bar{w})$. Our investigation focuses on deformed \cft{2}s at a finite temperature, and finite size deformed \cft{2}s at zero temperature, which are defined on appropriate cylinders. The expectation value of $\Theta$ vanishes on a cylinder and the $\Theta^2$ term in \cref{def_act} may be dropped from further consideration \cite{Chen:2018eqk}.

\subsection{Odd entanglement entropy}
\label{sn_oee}

We now focus our attention on a bipartite mixed state correlation measure termed the odd entanglement entropy (OEE), which approximately characterizes the von Neumann entropy for the partially transposed reduced density matrix of a given bipartite system \cite{Tamaoka:2018ned}. In this context we begin with a bipartite system comprising the subsystems $A$ and $B$, described by the reduced density matrix $\rho_{AB}$ defined on the Hilbert space $\mathcal{H}_{AB}=\mathcal{H}_A\otimes\mathcal{H}_B$, where $\mathcal{H}_A$ and $\mathcal{H}_B$ denote the Hilbert spaces for the subsystems $A$ and $B$ respectively. The partial transpose $\rho_{AB}^{T_B}$ for the reduced density matrix $\rho_{AB}$ with respect to the subsystem $B$ is then given by
\begin{align}
\label{pt_def}
\mel{e^{(A)}_ie^{(B)}_j}{\rho_{AB}^{T_B}}{e^{(A)}_ke^{(B)}_l}=\mel{e^{(A)}_ie^{(B)}_l}{\rho_{AB}}{e^{(A)}_ke^{(B)}_j},
\end{align}
where ${|e^{(A)}_i\rangle}$ and $|e^{(B)}_j\rangle$ describe orthonormal bases for the Hilbert spaces $\mathcal{H}_A$ and $\mathcal{H}_B$ respectively. The R{\'e}nyi odd entropy of order $n_o$ between the subsystems $A$ and $B$ may be defined as \cite{Kudler-Flam:2020url}
\begin{align}
\label{roe_def}
S_o^{(n_o)}\left(A:B\right)=\frac{1}{1-n_o}\log\left[\textrm{Tr}\left(\rho_{AB}^{T_B}\right)^{n_o}\right],
\end{align}
where $n_o$ is an odd integer. The OEE between the subsystems $A$ and $B$ may now be defined through the analytic continuation of the odd integer $n_o\to 1$ in \cref{roe_def} as follows \cite{Tamaoka:2018ned}
\begin{align}
\label{oee_def}
S_o(A:B)=\lim_{n_o\to 1}[S_{o}^{(n_o)}(A:B)]=\lim_{n_o\to 1}\frac{1}{1-n_o}\log\left[\textrm{Tr}\left(\rho_{AB}^{T_B}\right)^{n_o}\right].
\end{align}

\subsection{Odd entanglement entropy in a \cft{2}}
\label{sn_oee_cft}

The subsystems $A$ and $B$ in a \cft{2} may be characterized by the disjoint spatial intervals $[z_1,z_2]$ and $[z_3,z_4]$ in the complex plane [with $x_1<x_2<x_3<x_4\,,\, x={\rm Re}(z)$]. In \cite{Tamaoka:2018ned} the author advanced a replica technique to compute the OEE for bipartite systems in a \cft{2}. The replica construction involves an $n_o$ sheeted Riemann surface $\mathcal{M}_{n_o}$ (where $n_o\in 2\mathbb{Z}^+-1$) prepared through the cyclic and anti cyclic sewing of the branch cuts of $n_o$ copies of the original manifold $\mathcal{M}$ along the subsystems $A$ and $B$ respectively. Utilizing the replica technique, the trace of the partial transpose in \cref{roe_def} may be expressed in terms of the partition function on the $n_o$ sheeted replica manifold as follows \cite{Calabrese:2012ew,Calabrese:2012nk}
\begin{align}
\label{pt_ptfn}
\textrm{Tr}\left(\rho_{AB}^{T_B}\right)^{n_o}
=\frac{\mathbb{Z}\left[\mathcal{M}_{n_o}\right]}{\left(\mathbb{Z}\left[\mathcal{M}\right]\right)^{n_o}}\,.
\end{align}
The relation in \cref{pt_ptfn} may be utilized along with \cref{oee_def} to express the OEE in terms of the partition functions as follows
\begin{align}
\label{oee_ptfn}
S_o(A:B)=\lim_{n_o\to 1}\frac{1}{1-n_o}\log\left[\frac{\mathbb{Z}\left[\mathcal{M}_{n_o}\right]}{\left(\mathbb{Z}\left[\mathcal{M}\right]\right)^{n_o}}\right].
\end{align}
The partition function in \cref{pt_ptfn} may be expressed in terms of an appropriate four point correlation function of the twist and anti twist operators $\sigma_{n_o}$ and $\bar{\sigma}_{n_o}$ located at the end points of the subsystems $A$ and $B$ as follows \cite{Calabrese:2012ew,Calabrese:2012nk}
\begin{align}
\label{ptfn_twst}
\frac{\mathbb{Z}\left[\mathcal{M}_{n_o}\right]}{\left(\mathbb{Z}\left[\mathcal{M}\right]\right)^{n_o}}
=\left\langle \sigma_{n_o}(z_1,\bar{z}_{1})\bar{\sigma}_{n_o}(z_2,\bar{z}_{2})
\bar{\sigma}_{n_o}(z_3,\bar{z}_{3})\sigma_{n_o}(z_4,\bar{z}_{4}) \right\rangle.
\end{align}
We are now in a position to express the OEE between the subsystems $A$ and $B$ in terms of the four point twist correlator by combining \cref{roe_def,oee_def,pt_ptfn,ptfn_twst} as follows \cite{Tamaoka:2018ned,Calabrese:2012ew,Calabrese:2012nk}
\begin{align}
\label{oee_twst}
S_o(A:B)=\lim_{n_o\to 1}\frac{1}{1-n_o}\log\left[
\left\langle \sigma_{n_o}(z_1,\bar{z}_{1})\bar{\sigma}_{n_o}(z_2,\bar{z}_{2})
\bar{\sigma}_{n_o}(z_3,\bar{z}_{3})\sigma_{n_o}(z_4,\bar{z}_{4}) \right\rangle \right].
\end{align}
Note that $\sigma_{n_o}$ and $\bar{\sigma}_{n_o}$ represent primary operators in \cft{2} with the following conformal dimensions \cite{Calabrese:2012ew,Calabrese:2012nk,Calabrese:2014yza}
\begin{align}
\label{sigma_dim}
h_{n_o}=\bar{h}_{n_o}=\frac{c}{24}\left(n_o-\frac{1}{n_o}\right).
\end{align}
We also note in passing the conformal dimensions of the twist operators $\sigma_{n_o}^{2}$ and $\bar{\sigma}_{n_o}^{2}$, which are given as follows \cite{Calabrese:2012ew,Calabrese:2012nk,Calabrese:2014yza,Tamaoka:2018ned}
\begin{align}
\label{sigma_sq_dim}
h_{n_o}^{(2)}=\bar{h}_{n_o}^{(2)}=h_{n_{o}}=\frac{c}{24}\left(n_o-\frac{1}{n_o}\right).
\end{align}

\subsection{Holographic odd entanglement entropy}
\label{sn_hoee_rev}

We now follow \cite{Tamaoka:2018ned,KumarBasak:2021lwm} to present a brief review of the EWCS. Let $M$ be any specific time slice of a bulk static \ads{} geometry in the context of \ads{d+1}/\cft{d} framework. Consider a region $A$ in $\partial M$. The entanglement wedge of $A$ is given by the bulk region bounded by $A\cup\Gamma_A^{\rm min}$, where $\Gamma_A^{\rm min}$ is the RT surface for $A$. It has been proposed to be dual to the reduced density matrix $\rho_A$ \cite{Czech:2012bh,Wall:2012uf,Headrick:2014cta}. To define the EWCS, we subdivide $A=A_1\cup A_2$. A cross section of the entanglement wedge for $A_1\cup A_2$, denoted by $\Sigma_{A_1A_2}$, is defined such that it divides the wedge into two parts containing $A$ and $B$ separately. The EWCS between the subsystems $A_1$ and $A_2$ may then be defined as \cite{Takayanagi:2017knl}
\begin{align}
E_W (A_1:A_2)=\frac{\text{Area}\left(\Sigma_{A_1A_2}^{\rm min}\right)}{4G_N}\,,
\end{align}
where $\Sigma_{A_1A_2}^{\rm min}$ represents the minimal cross section of the entanglement wedge.

In \cite{Tamaoka:2018ned} the author proposed a holographic duality describing the difference of the OEE and the EE in terms of the bulk EWCS of the bipartite state in question as follows
\begin{align}
\label{hoee_def}
S_o (A_1:A_2) - S (A_1 \cup A_2) = E_W (A_1:A_2)\, ,
\end{align}
where $S(A_1 \cup A_2)$ is the EE for the subsystem $A_1 \cup A_2$, and $E_W (A_1:A_2)$ represents the EWCS between the subsystems $A_1$ and $A_2$ respectively.

\section{OEE in a \ttbar{} deformed \cft{2}}
\label{sn_oee_tt}

In this section we develop an appropriate replica technique similar to those described in \cite{Chen:2018eqk,Jeong:2019ylz,Basu:2023bov} for the computation of the OEE for various bipartite mixed state configurations in a \ttbar{} deformed \cft{2}. To this end we consider two spatial intervals $A$ and $B$ in a \ttbar{} deformed \cft{2} defined on a manifold $\mathcal{M}$. The partition functions on $\mathcal{M}$ and $\mathcal{M}_{n_o}$ for this deformed theory may be expressed in the path integral representation as follows [refer to \cref{def_act}]
\begin{align}
\label{ptfn_def}
\mathbb{Z}\left[\mathcal{M}\right] = \int_{\mathcal{M}} \mathcal{D}\phi\; e^{-\mathcal{I}_\text{QFT}^{(\mu)}[\phi]}\;,\qquad
\mathbb{Z}\left[\mathcal{M}_{n_o}\right] 
= \int_{\mathcal{M}_{n_o}} \mathcal{D}\phi\; e^{-\mathcal{I}_\text{QFT}^{(\mu)}[\phi]}\;.
\end{align}
When the deformation parameter $\mu$ is small, \cref{def_act,oee_ptfn,ptfn_def} may be utilized to express the OEE as
\begin{align}
\label{doee_act}
S_o^{(\mu)}(A:B)=\lim_{n_o\to 1}\frac{1}{1-n_o}\log\left[\frac{\int_{\mathcal{M}_{n_o}} 
\mathcal{D}\phi\; e^{-\mathcal{I}_\text{CFT}-\mu \int_{\mathcal{M}_{n_o}}(T\bar{T})}}{\left(\int_{\mathcal{M}}
\mathcal{D}\phi\; e^{-\mathcal{I}_\text{CFT}-\mu \int_{\mathcal{M}}(T\bar{T})}\right)^{n_o}}\right]\,,
\end{align}
where the superscript $\mu$ has been used to specify the OEE in the deformed \cft{2}. The exponential factors in \cref{doee_act} may be further expanded for small $\mu$ to arrive at
\begin{align}
S_o^{(\mu)}(A:B)&=\lim_{n_o\to 1}\frac{1}{1-n_o}\log\left[\frac{\int_{\mathcal{M}_{n_o}} \mathcal{D}\phi\; e^{-\mathcal{I}_\text{CFT}}\left(1-\mu \int_{\mathcal{M}_{n_o}}(T\bar{T})+\mathcal{O}(\mu^2)\right)}{\left[\int_{\mathcal{M}}\mathcal{D}\phi\; e^{-\mathcal{I}_\text{CFT}}\left(1-\mu \int_{\mathcal{M}}(T\bar{T})+\mathcal{O}(\mu^2)\right)\right]^{n_o}}\right] \nonumber \\
&=S_o^{(\text{CFT})}(A:B)+\lim_{n_o\to 1}\frac{1}{1-n_o}\log\left[\frac{\left(1-\mu \int_{\mathcal{M}_{n_o}}\Braket{T\bar{T}}_{\mathcal{M}_{n_o}}\right)}{\left(1-\mu\int_{\mathcal{M}}\Braket{T\bar{T}}_{\mathcal{M}}\right)^{n_o}}\right]\,. \label{doee_exp}
\end{align}
The term $S_o^{(\text{CFT})}(A:B)\equiv S_o^{(\mu=0)}(A:B)$ in \cref{doee_exp} represents the corresponding OEE for the undeformed \cft{2}. The expectation values of the \ttbar{} operator on the manifolds $\mathcal{M}$ and $\mathcal{M}_{n_o}$ appearing in \cref{doee_exp} are defined as follows
\begin{align}
\Braket{T\bar{T}}_{\mathcal{M}}=\frac{\int_{\mathcal{M}}\mathcal{D}\phi\;e^{-\mathcal{I}_\text{CFT}}(T\bar{T})}{\int_{\mathcal{M}}\mathcal{D}\phi\; e^{-\mathcal{I}_\text{CFT}}}\;,\qquad \Braket{T\bar{T}}_{\mathcal{M}_{n_o}}=\frac{\int_{\mathcal{M}_{n_o}}\mathcal{D}\phi\;e^{-\mathcal{I}_\text{CFT}}(T\bar{T})}{\int_{\mathcal{M}_{n_o}}\mathcal{D}\phi\; e^{-\mathcal{I}_\text{CFT}}}\;.
\end{align}
The second term on the right hand side of \cref{doee_exp} may be simplified to obtain the first order correction in $\mu$ to the OEE due to the \ttbar{} deformation as follows
\begin{align}
\label{coee_def}
\delta S_o(A:B) = -\mu\lim_{n_o\to 1}\frac{1}{1-n_o}\left[\int_{\mathcal{M}_{n_o}}\Braket{T\bar{T}}_{\mathcal{M}_{n_o}}-n_o \int_{\mathcal{M}}\Braket{T\bar{T}}_{\mathcal{M}}\right]\,.
\end{align}

\section{\ttbar{} deformed thermal \cft{2} and holography}
\label{sn_ft}

\subsection{OEE in a \ttbar{} deformed thermal \cft{2}}
\label{sn_oee_ft}

We now investigate the behavior of the deformed \cft{2} at a finite temperature $1/\beta$. The corresponding manifold $\mathcal{M}$ for this configuration is given by an infinitely long cylinder of circumference $\beta$ with the Euclidean time direction compactified by the periodic identification $\tau\sim\tau+\beta$. This cylindrical manifold $\mathcal{M}$ may be described by the complex coordinates \cite{Calabrese:2014yza}
\begin{align}
\label{cyl_coord}
w=x+i\tau \;, \qquad\qquad\qquad \bar{w}=x-i\tau \;,
\end{align}
with the spatial coordinate $x\in (-\infty,\infty)$ and the time coordinate $\tau\in (0,\beta)$. The cylinder $\mathcal{M}$ may be further expressed in terms of the complex plane $\mathbb{C}$ through the following conformal map \cite{Calabrese:2014yza}
\begin{align}
\label{map}
z=e^{\frac{2\pi w}{\beta}}\;, \qquad\qquad \bar{z}=e^{\frac{2\pi\bar{w}}{\beta}}\;,
\end{align}
where $(z, \bar{z})$ represent the coordinates on the complex plane. The transformation of the stress tensors under the conformal map described in \cref{map} is given as
\begin{align}
\label{stress_trfn}
T(w)=\left(\frac{2\pi z}{\beta}\right)^2T(z)-\frac{\pi^2c}{6\beta^2}\;,\qquad\qquad\bar{T}(\bar{w})=\left(\frac{2\pi \bar{z}}{\beta}\right)^2\bar{T}(\bar{z})-\frac{\pi^2c}{6\beta^2}\;.
\end{align}
The relations in \cref{stress_trfn} may be utilized to arrive at
\begin{align}
\label{tt_sg_sht}
\Braket{T(w)\bar{T}(\bar{w})}_{\mathcal{M}}=\left(\frac{\pi^2c}{6\beta^2}\right)^2,
\end{align}
where we have used the fact that $\Braket{T(z)}_{\mathbb{C}}=\Braket{\bar{T}(\bar{z})}_{\mathbb{C}}=0$ for the vacuum state of an undeformed \cft{2} described by the complex plane.
In the following subsections, we utilize \cref{coee_def} to compute the first order correction in $\mu$ to the OEE in a finite temperature \ttbar{} deformed \cft{2} for two disjoint intervals, two adjacent intervals and a single interval.

\subsubsection{Two disjoint intervals}
\label{sn_dj_ft}

We begin with the bipartite mixed state configuration of two disjoint spatial intervals $A=[x_1,x_2]$ and $B=[x_3,x_4]$ in a \ttbar{} deformed \cft{2} at a finite temperature $1/\beta$, defined on the cylindrical manifold $\mathcal{M}$ ($x_1<x_2<x_3<x_4$). Note that the intervals may also be represented as $A=[w_1,w_2]$ and $B=[w_3,w_4]$ with $\tau=0$ [cf. \cref{cyl_coord}]. The value of $\Braket{T\bar{T}}_{\mathcal{M}_{n_o}}$ on the replica manifold $\mathcal{M}_{n_o}$ may be computed by insertion of the \ttbar{} operator into the appropriate four point twist correlator as follows \cite{Calabrese:2004eu,Calabrese:2009qy}
\begin{align}
\label{tt_dj}
\int_{\mathcal{M}_{n_o}} \Braket{T\bar{T}}_{\mathcal{M}_{n_o}}
&= \sum_{k=1}^{n_o} \int_{\mathcal{M}} \frac{\Braket{T_{k}(w)\bar{T}_{k}(\bar{w})\sigma_{n_o}(w_{1}, \bar{w}_{1})\bar{\sigma}_{n_o}(w_{2}, \bar{w}_{2})\bar{\sigma}_{n_o}(w_{3}, \bar{w}_{3})\sigma_{n_o}(w_{4}, \bar{w}_{4})}_\mathcal{M}}{\Braket{\sigma_{n_o}(w_{1}, \bar{w}_{1})\bar{\sigma}_{n_o}(w_{2}, \bar{w}_{2})\bar{\sigma}_{n_o}(w_{3}, \bar{w}_{3})\sigma_{n_o}(w_{4}, \bar{w}_{4})}_\mathcal{M}} \\
&= \int_{\mathcal{M}}\frac{1}{n_o} \frac{\Braket{T^{(n_o)}(w)\bar{T}^{(n_o)}(\bar{w})\sigma_{n_o}(w_{1}, \bar{w}_{1})\bar{\sigma}_{n_o}(w_{2}, \bar{w}_{2})\bar{\sigma}_{n_o}(w_{3}, \bar{w}_{3})\sigma_{n_0}(w_{4}, \bar{w}_{4})}_\mathcal{M}}{\Braket{\sigma_{n_o}(w_{1}, \bar{w}_{1})\bar{\sigma}_{n_o}(w_{2}, \bar{w}_{2})\bar{\sigma}_{n_o}(w_{3}, \bar{w}_{3})\sigma_{n_o}(w_{4}, \bar{w}_{4})}_\mathcal{M}}\;. \nonumber
\end{align}
Here $T_k(w),\bar{T}_{k}(\bar{w})$ are the stress tensors of the undeformed \cft{2} on the $k^{th}$ sheet of the Riemann surface $\mathcal{M}_{n_o}$, while $T^{(n_o)}(w),\bar{T}^{(n_o)}(\bar{w})$ represent the stress tensors on $\mathcal{M}_{n_o}$ \cite{Calabrese:2004eu, Calabrese:2009qy}. $\sigma_{n_o}(w_{i},\bar{w}_{i}),\bar{\sigma}_{n_o}(w_{i},\bar{w}_{i})$ represent the twist operators located at the end points $w_i$ of the intervals. An identity described in \cite{Jeong:2019ylz} has been used to derive the last line of \cref{tt_dj}. The relation in \cref{stress_trfn} may now be utilized to transform the stress tensors from the cylindrical manifold to the complex plane. The following Ward identities are then employed to express the correlation functions involving the stress tensors in terms of the twist correlators on the complex plane
\begin{align}
\label{ward_hol}
& \Braket{T^{(n_o)}(z)\mathcal{O}_{1}(z_{1},\bar{z}_{1})\ldots \mathcal{O}_{m}(z_{m},\bar{z}_{m})}_{\mathbb{C}} \\
& = \sum_{j=1}^{m} \left(\frac{h_j}{(z-z_j)^2}+\frac{1}{(z-z_j)}\partial_{z_j}\right) \Braket{\mathcal{O}_{1}(z_{1},\bar{z}_{1})\ldots \mathcal{O}_{m}(z_{m},\bar{z}_{m})}_{\mathbb{C}}\,, \nonumber \\
& \Braket{\bar{T}^{(n_o)}(\bar{z})\mathcal{O}_{1}(z_{1},\bar{z}_{1})\ldots \mathcal{O}_{m}(z_{m},\bar{z}_{m})}_{\mathbb{C}} \nonumber \\
& = \sum_{j=1}^{m}\left(\frac{\bar{h}_j}{(\bar{z}-\bar{z}_j)^2}+\frac{1}{(\bar{z}-\bar{z}_j)}\partial_{\bar{z}_j}\right)
\Braket{\mathcal{O}_{1}(z_{1},\bar{z}_{1})\ldots \mathcal{O}_{m}(z_{m},\bar{z}_{m})}_{\mathbb{C}}\,, \nonumber
\end{align}
where $\mathcal{O}_{i}$s represent arbitrary primary operators with conformal dimensions $(h_i ,\bar{h}_i)$.
Utilizing \cref{stress_trfn}, we may now express the expectation value in \cref{tt_dj} as
\begin{align}
\label{tt_dj_2}
\int_{\mathcal{M}_{n_o}} \Braket{T\bar{T}}_{\mathcal{M}_{n_o}}
& = \frac{1}{n_o}\int_{\mathcal{M}}\frac{1}{\Braket{\sigma_{n_o}(z_{1}, \bar{z}_{1})\bar{\sigma}_{n_o}(z_{2}, \bar{z}_{2})\bar{\sigma}_{n_o}(z_{3}, \bar{z}_{3})\sigma_{n_o}(z_{4}, \bar{z}_{4})}_{\mathbb{C}}} \\
& \times \left[ -\frac{\pi^2 c\,n_o}{6 \beta^2}+\left(\frac{2 \pi z}{\beta} \right)^2 \sum_{j=1}^{4} \left(\frac{h_j}{(z-z_j)^2}+\frac{1}{(z-z_j)}\partial_{z_j}\right) \right] \nonumber \\
& \times \left[ -\frac{\pi^2 c\,n_o}{6 \beta^2}+\left(\frac{2 \pi \bar{z}}{\beta} \right)^2 \sum_{k=1}^{4} \left(\frac{\bar{h}_k}{(\bar{z}-\bar{z}_k)^2}+\frac{1}{(\bar{z}-\bar{z}_k)}\partial_{\bar{z}_k}\right) \right] \nonumber \\
& \times \Braket{\sigma_{n_o}(z_{1}, \bar{z}_{1})\bar{\sigma}_{n_o}(z_{2}, \bar{z}_{2})\bar{\sigma}_{n_o}(z_{3}, \bar{z}_{3})\sigma_{n_o}(z_{4}, \bar{z}_{4})}_{\mathbb{C}}\, , \nonumber
\end{align}
where $h_i=\bar{h}_i=h_{n_o} (i=1,2,3,4)$ [see \cref{sigma_dim}]. The four point twist correlator in \cref{tt_dj_2} for two disjoint intervals in proximity described by the $t$-channel is given by \cite{Tamaoka:2018ned,Fitzpatrick:2014vua}
\begin{align}
\label{4pt_fn}
& \Braket{\sigma_{n_o}(z_{1},\bar{z}_{1})\bar{\sigma}_{n_o}(z_{2},\bar{z}_{2})
\bar{\sigma}_{n_o}(z_{3},\bar{z}_{3})\sigma_{n_o}(z_{4},\bar{z}_{4})}_{\mathbb{C}}
\approx \abs{z_{14}z_{23}}^{-4 h_{n_o}}
\left(\frac{1+\sqrt{\eta}}{1-\sqrt{\eta}}\right)^{-h_{n_o}^{(2)}}
\left(\frac{1+\sqrt{\bar{\eta}}}{1-\sqrt{\bar{\eta}}}\right)^{-\bar{h}_{n_o}^{(2)}}.
\end{align}
The conformal dimensions $h_{n_o}$, $h_{n_o}^{(2)}$ and $\bar{h}_{n_o}^{(2)}$ in \cref{4pt_fn} are given in \cref{sigma_dim,sigma_sq_dim}. We have defined the cross ratio $\eta:=\frac{z_{12} z_{34}}{z_{13} z_{24}}$ where $z_{ij}\equiv z_i-z_j$.

We are now in a position to obtain the first order correction due to $\mu$ in the OEE of two disjoint intervals in a \ttbar{} deformed finite temperature \cft{2} by substituting \cref{tt_sg_sht,tt_dj_2,4pt_fn} into \cref{coee_def} as follows
\begin{align}
\label{coee_dj_int}
\delta S_{o}(A:B) = -\frac{\mu c^2 \pi ^4 \sqrt{\eta}}{18\beta^4 z_{21} z_{32} z_{41} z_{43}}\int_{\mathcal{M}} z^2 &\left[\frac{z_{32} z_{42} [z_{31} (2z-3z_1+z_4)\sqrt{\eta}+z_{43} (z-z_1)]}{(z-z_1)^2} \right. \\
& +\frac{z_{31} z_{41} [z_{42} (2z-3z_2+z_3)\sqrt{\eta}-z_{43} (z-z_2)]}{(z-z_2)^2} \nonumber \\
& -\frac{z_{42} z_{41} [z_{31}(2z+z_2-3z_3) \sqrt{\eta}-z_{21}(z-z_3)]}{(z-z_3)^2} \nonumber \\
& \left. -\frac{z_{31} z_{32} [z_{42} (2z+z_1-3z_4)\sqrt{\eta}+z_{21} (z-z_4)]}{(z-z_4)^2}\right]+h.c.\nonumber
\end{align}
The detailed derivation of the definite integrals in \cref{coee_dj_int} has been provided in \cref{app_dj}. These results may be used to arrive at
\begin{align}
\label{coee_dj_z}
\delta S_{o} (A:B) & = \frac{\mu c^2 \pi^3 }{36 \beta^2} \left[ 
\frac{ \left\lbrace \left( \sqrt{\frac{z_{42} z_{43}}{z_{21} z_{31}}}+1 \right) z_1+z_4 \right\rbrace }{ z_{41} } 
\log \left[ \frac{z_1}{z_2} \right] \right. \\
& \left. + \frac{\left(\sqrt{\frac{z_{21} z_{43}}{z_{31} z_{42}}}-2\right) \left(z_1 z_2-z_3 z_4\right) }{z_{32} z_{41}} 
\log \left[ \frac{z_2}{z_3} \right] 
+ \frac{\left\lbrace z_1 - \left(\sqrt{\frac{z_{21} z_{31}}{z_{42} z_{43}}}-1 \right) z_4 \right\rbrace}{ z_{41} } 
\log \left[ \frac{z_3}{z_4} \right] + h.c. \right]. \nonumber
\end{align}
We may now substitute $z_i = \bar{z}_i = e^{\frac{2\pi x_i}{\beta}}$ (at $\tau_{i}=0$) into \cref{coee_dj_z} to finally obtain the leading order corrections to the OEE as follows
\begin{align}
\label{coee_dj_x}
\delta S_{o} (A:B) & = -\frac{\mu c^2 \pi^4}{9 \beta ^3}
\sqrt{\frac{\sinh \left(\frac{\pi  x_{21}}{\beta }\right) \sinh \left(\frac{\pi  x_{43}}{\beta }\right)}{\sinh \left(\frac{\pi  x_{31}}{\beta }\right) \sinh \left(\frac{\pi  x_{42}}{\beta }\right)}}\left[x_{21} \coth \left(\frac{\pi  x_{21}}{\beta }\right)\right. \\
& \left. -x_{32} \coth \left(\frac{\pi  x_{32}}{\beta }\right) 
- x_{41} \coth \left(\frac{\pi  x_{41}}{\beta }\right)+x_{43} 
\coth \left(\frac{\pi  x_{43}}{\beta }\right)\right] \nonumber \\
& -\frac{\mu c^2 \pi^4}{9 \beta ^3}\left[x_{32} \coth \left(\frac{\pi  x_{32}}{\beta }\right)+x_{41} \coth \left(\frac{\pi  x_{41}}{\beta }\right)\right],\nonumber
\end{align}
where $x_{ij}\equiv x_i-x_j$. It is worth noting that the last term in the above expression is nothing but the leading order corrections to the entanglement entropy of the two disjoint intervals in the $t$-channel. Remarkably, in the low temperature limit $\beta\gg x_{ij}$, the corrections to the OEE scales exactly like that for the entanglement entropy, $\frac{-2\mu \pi^3c^2}{9\beta^2}$ \cite{Chen:2018eqk}. In particular, in the zero temperature limit $\beta\to\infty$, the corrections vanish conforming to our expectations.

\subsubsection{Two adjacent intervals}
\label{sn_adj_ft}

We now turn our attention to the bipartite mixed state configuration of two adjacent intervals $A=[x_1,x_2]$ and $B=[x_2,x_3]$
in a \ttbar{} deformed \cft{2} at a finite temperature $1/\beta$ ($x_1<x_2<x_3$). As earlier the intervals may be expressed as $A=[w_1,w_2]$ and $B=[w_2,w_3]$ with $\tau=0$. The value of $\Braket{T\bar{T}}_{\mathcal{M}_{n_o}}$ for two adjacent intervals may be evaluated in a manner similar to that of two disjoint intervals as follows
\begin{align}
\label{tt_adj}
\int_{\mathcal{M}_{n_o}} \Braket{T\bar{T}}_{\mathcal{M}_{n_o}} &= \int_{\mathcal{M}}\frac{1}{n_o} \frac{\Braket{T^{(n_o)}(w)\bar{T}^{(n_o)}(\bar{w})\sigma_{n_o}(w_{1}, \bar{w}_{1})\bar{\sigma}^{2}_{n_o}(w_{2}, \bar{w}_{2})\sigma_{n_o}(w_{3}, \bar{w}_{3})}_{\mathcal{M}}}{\Braket{\sigma_{n_o}(w_{1}, \bar{w}_{1})\bar{\sigma}^{2}_{n_o}(w_{2}, \bar{w}_{2})\sigma_{n_o}(w_{3}, \bar{w}_{3})}_{\mathcal{M}}}\;.
\end{align}
As before the relations in \cref{stress_trfn,ward_hol} may be utilized to express the expectation value in \cref{tt_adj} as follows
\begin{align}
\label{tt_adj_2}
\int_{\mathcal{M}_{n_o}} \Braket{T\bar{T}}_{\mathcal{M}_{n_o}} & = \frac{1}{n_o}\int_{\mathcal{M}}\frac{1}{\Braket{\sigma_{n_o}(z_{1}, \bar{z}_{1})\bar{\sigma}^{2}_{n_o}(z_{2}, \bar{z}_{2})\sigma_{n_o}(z_{3}, \bar{z}_{3})}_{\mathbb{C}}} \\
& \times \left[ -\frac{\pi^2 c\,n_o}{6 \beta^2}+\left(\frac{2 \pi z}{\beta} \right)^2 \sum_{j=1}^{3} \left(\frac{h_j}{(z-z_j)^2}+\frac{1}{(z-z_j)}\partial_{z_j}\right) \right] \nonumber \\
& \times \left[ -\frac{\pi^2 c\,n_o}{6 \beta^2}+\left(\frac{2 \pi \bar{z}}{\beta} \right)^2 \sum_{k=1}^{3} \left(\frac{\bar{h}_k}{(\bar{z}-\bar{z}_k)^2}+\frac{1}{(\bar{z}-\bar{z}_k)}\partial_{\bar{z}_k}\right) \right] \nonumber \\
& \times \Braket{\sigma_{n_o}(z_{1}, \bar{z}_{1})\bar{\sigma}^{2}_{n_o}(z_{2}, \bar{z}_{2})\sigma_{n_o}(z_{3}, \bar{z}_{3})}_{\mathbb{C}}\,. \nonumber
\end{align}
In \cref{tt_adj_2} we have $h_1=h_3=h_{n_o},h_2=h^{(2)}_{n_o}$ with $\bar{h}_{i}=h_{i} \; (i=1,2,3)$ [see \cref{sigma_dim,sigma_sq_dim}]. The three point twist correlator in \cref{tt_adj_2} is given by \cite{francesco2012conformal}
\begin{align}
\label{3pt_fn}
\Braket{\sigma_{n_o}(z_{1}, \bar{z}_{1})\bar{\sigma}^{2}_{n_o}(z_{2}, \bar{z}_{2})\sigma_{n_o}(z_{3}, \bar{z}_{3})}_{\mathbb{C}}
=\frac{\mathcal{C}_{\sigma_{n_o}\bar{\sigma}_{n_o}^{2}\sigma_{n_o}}}
{\left( z^{h^{(2)}_{n_o}}_{12}z^{h^{(2)}_{n_o}}_{23}z^{2h_{n_o}-h^{(2)}_{n_o}}_{13} \right)
\left( \bar{z}^{\bar{h}^{(2)}_{n_o}}_{12}\bar{z}^{\bar{h}^{(2)}_{n_o}}_{23}\bar{z}^{2\bar{h}_{n_o}-\bar{h}^{(2)}_{n_o}}_{13} \right) } \;,
\end{align}
where $\mathcal{C}_{\sigma_{n_e}\bar{\sigma}_{n_e}^{2}{\sigma}_{n_e}}$ is the relevant OPE coefficient. The first order correction due to $\mu$ in the OEE of two adjacent intervals in a \ttbar{} deformed thermal \cft{2} may now be obtained by substituting \cref{tt_sg_sht,tt_adj_2,3pt_fn} into \cref{coee_def} as follows
\begin{align}
\label{coee_adj_int}
\delta S_{o} (A:B) = -\frac{\mu c^2\pi^4}{18 \beta^4}
\int_{\mathcal{M}}z^2\Bigg[
\frac{1}{\left(z-z_1\right)^2}&+\frac{1}{\left(z-z_2\right)^2}+\frac{1}{\left(z-z_3\right)^2} \\
& +\frac{\left(-3 z+z_1+z_2+z_3\right) }{\left(z-z_1\right) \left(z-z_2\right) \left(z-z_3\right)}
+ h.c. \Bigg]. \nonumber
\end{align}
The technical details of the definite integrals in \cref{coee_adj_int} have been included in \cref{app_adj}. The correction to the OEE may then be expressed as
\begin{align}
\label{coee_adj_z}
\delta S_{o} (A:B)
&= -\frac{\mu c^2\pi^3}{36\beta^2} \left[\frac{\left(z_1^2-z_2 z_3\right) \log \left(\frac{z_1}{z_2}\right)}{z_{12} z_{13}}+\frac{\left(z_1 z_2-z_3^2\right) \log \left(\frac{z_2}{z_3}\right)}{z_{23}z_{13}} + h.c. \right] .
\end{align}
As earlier we may now restore the $x$ coordinates by inserting $z_i = \bar{z}_i = e^{\frac{2\pi x_i}{\beta}}$ (at $\tau_{i}=0$) into \cref{coee_adj_z} to arrive at
\begin{align}
\label{coee_adj_x}
\delta S_{o} (A:B)
&= - \left(\frac{\mu c^2\pi^4}{36\beta^3}\right) \frac{x_{21} \cosh \left(\frac{2 \pi  x_{21}}{\beta }\right)+x_{32} \cosh \left(\frac{2 \pi  x_{32}}{\beta }\right) - x_{31} \cosh \left(\frac{2 \pi  x_{31}}{\beta }\right) }{\sinh \left(\frac{\pi  x_{21}}{\beta }\right) \sinh \left(\frac{\pi  x_{32}}{\beta }\right) \sinh \left(\frac{\pi  x_{31}}{\beta }\right)}
\end{align}
Once again, we see that the leading order corrections to the OEE scales exactly like that of the entanglement entropy in the low temperature limit $\beta\gg x_{ij}$. It is interesting to note that we are unable to reproduce the above result by taking an appropriate adjacent limit of the corrections to the disjoint intervals given in \cref{coee_dj_x}. However, this does not lead to any contradiction since our field theory results are perturbative and there is no a priori reason to believe that a limiting analysis holds in each order of conformal perturbation theory. More evidence towards this mismatch will be provided from a holographic viewpoint in \cref{sn_adj_ft_hol}.

\subsubsection{A single interval}
\label{sn_sg_ft}

We finally focus on the case of a single interval $A=[-\ell,0]$ in a thermal \ttbar{} deformed \cft{2} ($\ell>0$). To this end it is required to consider two auxiliary intervals $B_1=[-L, -\ell]$ and $B_2=[0,L]$ on either side of the interval $A$ with $B\equiv B_1\cup B_2$ ($L\gg\ell$) \cite{Calabrese:2014yza}. The intervals may be equivalently represented by the coordinates $B_1=[x_1,x_2]$, $A=[x_2,x_3]$ and $B_2=[x_3,x_4]$, with $x_1=-L,x_2=-\ell,x_3=0,x_4=L$ and $x_1<x_2<x_3<x_4$. As before the intervals may also be characterized as $B_1=[w_1,w_2]$, $A=[w_2,w_3]$ and $B_2=[w_3,w_4]$ with $\tau=0$. The OEE for the mixed state configuration of the single interval $A$ is then evaluated by implementing the bipartite limit $L\to\infty$ ($B_1\cup B_2\to A^c$) subsequent to the replica limit $n_o\to 1$ \cite{Calabrese:2014yza}. For the configuration described above, the integral of $\Braket{T\bar{T}}_{\mathcal{M}_{n_o}}$ on the replica manifold is given by
\begin{align}
\label{tt_sg}
\int_{\mathcal{M}_{n_o}} \Braket{T\bar{T}}_{\mathcal{M}_{n_o}} &= \int_{\mathcal{M}}\frac{1}{n_o} \frac{\Braket{T^{(n_o)}(w)\bar{T}^{(n_o)}(\bar{w})\sigma_{n_o}(w_{1}, \bar{w}_{1})\bar{\sigma}^{2}_{n_o}(w_{2}, \bar{w}_{2})\sigma^{2}_{n_o}(w_{3}, \bar{w}_{3})\bar{\sigma}_{n_o}(w_{4}, \bar{w}_{4})}}{\Braket{\sigma_{n_o}(w_{1}, \bar{w}_{1})\bar{\sigma}^{2}_{n_o}(w_{2}, \bar{w}_{2})\sigma^{2}_{n_o}(w_{3}, \bar{w}_{3})\bar{\sigma}_{n_o}(w_{4}, \bar{w}_{4})}}\;.
\end{align}
As earlier \cref{tt_sg} may be simplified by utilizing \cref{stress_trfn,ward_hol} as follows
\begin{align}
\label{tt_sg_2}
\int_{\mathcal{M}_{n_o}} \Braket{T\bar{T}}_{\mathcal{M}_{n_o}}
& = \frac{1}{n_o} \int_{\mathcal{M}} \frac{1}{\Braket{\sigma_{n_o}(z_{1}, \bar{z}_{1})\bar{\sigma}^{2}_{n_o}(z_{2}, \bar{z}_{2})\sigma^{2}_{n_o}(z_{3}, \bar{z}_{3})\bar{\sigma}_{n_o}(z_{4}, \bar{z}_{4})}} \\
& \times \left[ -\frac{\pi^2 c\,n_o}{6 \beta^2}+\left(\frac{2 \pi z}{\beta} \right)^2 \sum_{j=1}^{4} \left(\frac{h_j}{(z-z_j)^2}+\frac{1}{(z-z_j)}\partial_{z_j}\right) \right] \nonumber \\
& \times \left[ -\frac{\pi^2 c\,n_o}{6 \beta^2}+\left(\frac{2 \pi \bar{z}}{\beta} \right)^2 \sum_{k=1}^{4} \left(\frac{\bar{h}_k}{(\bar{z}-\bar{z}_k)^2}+\frac{1}{(\bar{z}-\bar{z}_k)}\partial_{\bar{z}_k}\right) \right] \nonumber \\
& \times\Braket{\sigma_{n_o}(z_{1}, \bar{z}_{1})\bar{\sigma}^{2}_{n_o}(z_{2}, \bar{z}_{2})\sigma^{2}_{n_o}(z_{3}, \bar{z}_{3})\bar{\sigma}_{n_o}(z_{4}, \bar{z}_{4})}_{\mathcal{C}} \,, \nonumber
\end{align}
where $h_1=h_4=h_{n_o},h_2=h_3=h^{(2)}_{n_o}$ with $\bar{h}_{i}=h_{i} \; (i=1,2,3,4)$ [see \cref{sigma_dim,sigma_sq_dim}]. The four point twist correlator in \cref{tt_sg_2} is given by \cite{Calabrese:2014yza}
\begin{align}
\label{4pt_fn_sg}
\Braket{\sigma_{n_o}(z_{1}, \bar{z}_{1})\bar{\sigma}^{2}_{n_o}(z_{2}, \bar{z}_{2})\sigma^{2}_{n_o}(z_{3}, \bar{z}_{3})\bar{\sigma}_{n_o}(z_{4}, \bar{z}_{4})}
= c_{n_o}c^{(2)}_{n_o}\left(\frac{\mathcal{F}_{n_o}(\eta)}{z^{2h_{n_o}}_{14} z^{2h^{(2)}_{n_o}}_{23} \eta^{h^{(2)}_{n_o}}}\right)  \left(\frac{\bar{\mathcal{F}}_{n_o}(\bar{\eta})}{\bar{z}^{2\bar{h}_{n_o}}_{14} \bar{z}^{2\bar{h}^{(2)}_{n_o}}_{23}\bar{\eta}^{\bar{h}^{(2)}_{n_o}}}\right) ,
\end{align}
where $c_{n_o}$ and $c_{n_o}^{(2)}$ are the normalization constants. The functions $\mathcal{F}_{n_o}(\eta)$ and $\bar{\mathcal{F}}_{n_o}(\bar{\eta})$ in \cref{4pt_fn_sg} satisfy the following OPE limits
\begin{align}
\mathcal{F}_{n_o}(1)\bar{\mathcal{F}}_{n_o}(1)=1 \; , \qquad \mathcal{F}_{n_o}(0)\bar{\mathcal{F}}_{n_o}(0)=\frac{ \mathcal{C}_{\sigma_{n_o}\bar{\sigma}_{n_o}^{2}\bar{\sigma}_{n_o}}}{c_{n_o}^{(2)}}\;, \nonumber
\end{align}
where $\mathcal{C}_{\sigma_{n_o}\bar{\sigma}_{n_o}^{2}\bar{\sigma}_{n_o}}$ represents the relevant OPE coefficient. As earlier \cref{tt_sg_sht,tt_sg_2,4pt_fn_sg} may be substituted into \cref{coee_def} to arrive at
\begin{align}
\label{coee_sg_int}
\delta S_o (A:B)=& -\frac{\mu c^2 \pi^4}{18 \beta^4} \int_{\mathcal{M}} 
\Bigg[ \sum_{j=1}^{4}\frac{z^2}{(z-z_j)^2}
-\sum_{j=1}^{4}\frac{z^2}{(z-z_j)}\partial_{z_j}
\bigg(\log\left[z^{2}_{23} z^{2}_{14}\ \eta f(\eta)\right]\bigg) + h.c. \Bigg].
\end{align}
The functions $f(\eta)$ and $\bar{f}(\bar{\eta})$ introduced in \cref{coee_sg_int} are defined as follows
\begin{align}
\lim_{n_o\to 1} [\mathcal{F}_{n_o} (\eta)]^{\frac{1}{1-n_o}} = [f(\eta)]^{c/12} \;\; ,
\qquad \lim_{n_o\to 1} [\bar{\mathcal{F}}_{n_o} (\bar{\eta})]^{\frac{1}{1-n_o}} = [\bar{f}(\bar{\eta})]^{c/12}\;. \nonumber
\end{align}
The first order correction due to $\mu$ in the OEE of a single interval in a \ttbar{} deformed \cft{2} at a finite temperature $1/\beta$ may now be computed from \cref{coee_sg_int} by reverting back to the coordinates involving $\ell, L$ and implementing the bipartite limit $L\to\infty$ as follows
\begin{align}
\label{coee_sg}
\delta S_o (A:A^c) & = -\frac{2 \mu c^2 \pi^4 \ell}{9\beta^3}
\left(\frac{1}{ e^{\frac{2 \pi \ell}{\beta}} -1 } - e^{-\frac{2 \pi \ell}{\beta}} \frac{ f' \left[ e^{-\frac{2 \pi \ell}{\beta}} \right]}{2 f \left[ e^{-\frac{2 \pi \ell}{\beta}} \right]}  \right) \\
& - \lim_{L\to\infty} \left[  \frac{\mu c^2 \pi ^4 L}{9\beta^3} 
\coth \left( \frac{2 \pi L}{\beta} \right) \right] . \nonumber
\end{align}
The technical details of the integrals necessary to arrive at \cref{coee_sg} from \cref{coee_sg_int} have been provided in \cref{app_sg}. Note that the second term on the right hand side of \cref{coee_sg} represents a divergent piece in the OEE for a single interval. Essentially, the quantity inside the parenthesis of the second term is the leading order correction to the entanglement entropy of the interval $A\cup B_1\cup B_2$. In the bipartite limit $L\to\infty$, this represents the entanglement entropy of the entire system and hence should be vanishing. The IR divergence is an artifact of placing a cutoff in a continuum field theory\footnote{Similar divergences are observed in the usual CFT$_2$ in its vacuum state. The entanglement entropy for a single interval of length $\ell$ at zero temperature is given by $\frac{c}{3}\log \left(\frac{\ell}{\epsilon}\right)$ \cite{Calabrese:2004eu}, which diverges logarithmically as $\ell\to \infty$.}. 

Interestingly the universal finite piece of the OEE for a single interval in a \ttbar-deformed CFT$_2$ may be rewritten up to leading order in the deformation as follows
\begin{align}
	S_o (A:A^c)=S_A-S_A^{\text{Th}}\,,
\end{align}
where the thermal entropy $S_A^{\text{Th}}$ is now given by
\begin{align}
	S_A^{\text{Th}}=\frac{\pi c \ell}{3\beta}\left(1-\mu\frac{\pi^3c}{3\beta^2}\right)\,.\label{Thermal-entropy}
\end{align}
A comparison of the above expression to the thermal contribution in the undeformed case, $\frac{\pi c \ell}{3\beta}$ \cite{Calabrese:2014yza}, indicates that the thermal entropy receives non-trivial corrections due to the \ttbar-deformation.

\subsection{Holographic OEE in a \ttbar{} deformed thermal \cft{2}}
\label{sn_hoee_ft}

We now turn our attention to the holographic description of the OEE as advanced in \cite{Tamaoka:2018ned} for various bipartite mixed states in a \ttbar{} deformed \cft{2} at a finite temperature $1/\beta$. The holographic dual of a \ttbar{} deformed \cft{2} is described by the bulk \ads{3} geometry corresponding to the undeformed \cft{2} with a finite cut-off radius $r_c$ given as follows \cite{McGough:2016lol}
\begin{align}
\label{cutoff_rad}
r_c=\sqrt{\frac{6 R^4}{\pi c\mu}}=\frac{R^2}{\epsilon}\,.
\end{align}
In \cref{cutoff_rad} $\mu$ is the deformation parameter, $c$ is the central charge, $\epsilon$ is the UV cut-off of the field theory, and $R$ is the \ads{3} radius. For a \ttbar{} deformed \cft{2} at a finite temperature $1/\beta$, the corresponding bulk dual is characterized by a BTZ black hole \cite{Banados:1998gg} with a finite cut-off, represented by \cite{McGough:2016lol}
\begin{align}
\label{btz_metric}
\mathrm{d}s^2=-\frac{r^2-r_h^2}{R^2}\mathrm{d}t^2+\frac{R^2}{r^2-r_h^2}\mathrm{d}r^2+r^2\mathrm{d}\tilde{x}^2\,.
\end{align}
In the above metric, the horizon of the black hole is located at $r=r_h$, with $\beta=\frac{2\pi R^2}{r_h}$ as the inverse temperature of the black hole and the dual \cft{2}. For simplicity from now onwards we set the \ads{} radius $R=1$. The metric on the \ttbar{} deformed \cft{2}, located at the cut-off radius $r=r_c$, is conformal to the bulk metric at $r=r_c$ as follows \cite{Chen:2018eqk, Jeong:2019ylz}
\begin{align}
\label{cft_metric}
\mathrm{d}s^2=-\mathrm{d}t^2+\frac{\mathrm{d}\tilde{x}^2}{1-\frac{r_h^2}{r_c^2}}
\equiv -\mathrm{d}t^2+\mathrm{d}x^2~~,~~
x=\frac{r_c\,\tilde{x}}{\sqrt{\displaystyle r_c^2-r_h^2}},
\end{align}
where $x$ represents the spatial coordinate on the deformed \cft{2}. To compute the EWCS, we embed the BTZ black hole described by \cref{btz_metric} in $\mathbb{R}^{2,2}$ as follows \cite{Kusuki:2019evw}
\begin{align}
\label{embed}
\mathrm{d}s^2
=\eta_{AB} \mathrm{d}X^A\mathrm{d}X^B =-\mathrm{d}X^2_0-\mathrm{d}X^2_1+\mathrm{d}X^2_2+\mathrm{d}X^2_3\:,
\qquad X^2=-1\,.
\end{align}
The metric in \cref{btz_metric} may then be described by these embedding coordinates as follows \cite{Carlip:1995qv,Shenker:2013pqa}
\begin{align}
\label{global_coord}
X_0(t,r,x)&=\sqrt{\frac{r^2}{{r_h}^2}-1} ~~ \sinh \left(\frac{2 \pi  t}{\beta }\right), \\
X_1(t,r,x)&=\frac{r}{r_h} \cosh \left(\frac{2 \pi  \tilde{x}}{\beta }\right), \nonumber\\
X_2(t,r,x)&=\sqrt{\frac{r^2}{{r_h}^2}-1} ~~ \cosh \left( \frac{2 \pi  t}{\beta } \right), \nonumber \\
X_3(t,r,x)&=\frac{r}{r_h} \sinh \left(\frac{2 \pi  \tilde{x}}{\beta }\right). \nonumber
\end{align}
Note that for convenience the embedding coordinates in \cref{global_coord} are parameterized in terms of the coordinate $x$ described in \cref{cft_metric}. We also introduce a new coordinate $u = 1/r$ to simplify later calculations, with $u_c \equiv 1/r_c$ and $u_h \equiv 1/r_h$. We also note the Brown Henneaux formula $G_N=3/(2c)$ described in \cite{Brown:1986nw}, which will be extensively used in later sections. In the following subsections we apply the methods described above to compute the holographic OEE from \cref{hoee_def} for two disjoint intervals, two adjacent intervals, and a single interval in a \ttbar{} deformed thermal holographic \cft{2}.

\subsubsection{Two disjoint intervals}
\label{sn_dj_ft_hol}

\begin{figure}[H]
\centering
\includegraphics[scale=0.50]{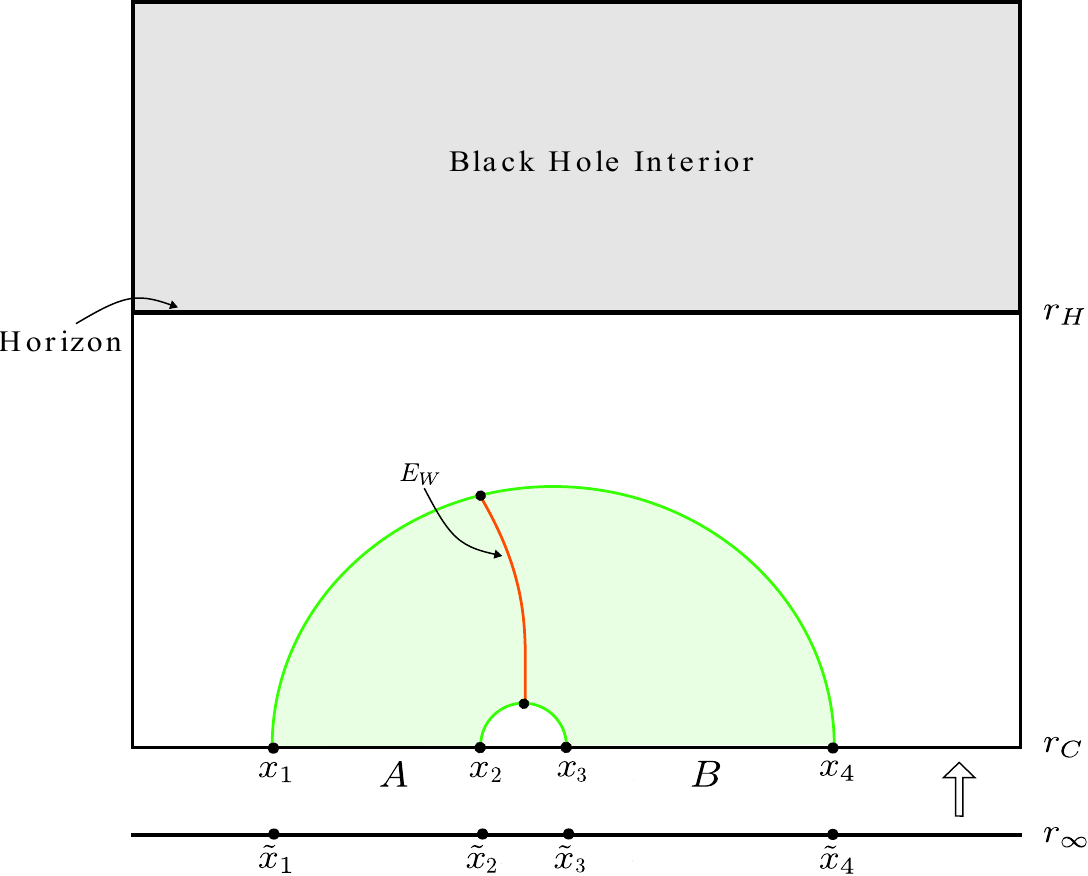}
\caption{EWCS for two disjoint intervals in a $T\bar{T}$ deformed \cft{2}. Figure based on \cite{Basu:2023bov}.}
\label{fg_EW_dj}
\end{figure}

We begin with the two disjoint spatial intervals $A=[x_1,x_2]$ and $B=[x_3,x_4]$ with $x_1<x_2<x_3<x_4$ as described in \cref{sn_dj_ft}. The setup has been shown in \cref{fg_EW_dj}. The EWCS involving the bulk points $X(s_1),X(s_2),X(s_3),X(s_4)$ is given by \cite{Kusuki:2019evw}
\begin{align}
\label{EW_dj_def}
E_W = \frac{1}{4G_N} \cosh ^{-1} \left(  \frac{1+\sqrt{u}}{\sqrt{v}}  \right),
\end{align}
where
\begin{align}
u=\dfrac{\xi^{-1}_{12}\xi^{-1}_{34}}{\xi^{-1}_{13}\xi^{-1}_{24}}\;,\qquad v=\dfrac{\xi^{-1}_{14}\xi^{-1}_{23}}{\xi^{-1}_{13}\xi^{-1}_{24}}\;,\qquad
\xi^{-1}_{ij}=-X(s_i)\cdot X(s_j)\;.
\end{align}
The four points on the boundary may be expressed in the global coordinates as $X(0,r_c,x_i)$ for $i=1,2,3,4$. The corresponding EWCS may then be computed from \cref{EW_dj_def} as

\begin{align}
\label{EW_dj}
&E_W(A:B)\\
&=\frac{1}{4G_N} \cosh ^{-1} \left(  \sqrt{\frac{ 
\left[ u_c^2-u_h^2+u_h^2 \cosh \left(  \frac{\sqrt{u_h^2-u_c^2}~x_{31} }{u_h^2} \right) \right]
\left[ u_c^2-u_h^2+u_h^2 \cosh \left(  \frac{\sqrt{u_h^2-u_c^2}~x_{42} }{u_h^2} \right) \right]  }{  
\left[ u_c^2-u_h^2+u_h^2 \cosh \left(  \frac{\sqrt{u_h^2-u_c^2}~x_{32} }{u_h^2} \right) \right]  
\left[ u_c^2-u_h^2+u_h^2 \cosh \left(  \frac{\sqrt{u_h^2-u_c^2}~x_{41} }{u_h^2} \right) \right]  }}  \right. \nonumber \\
& + \left. \sqrt{\frac{ 
\left[ u_c^2-u_h^2+u_h^2 \cosh \left(  \frac{\sqrt{u_h^2-u_c^2}~x_{21} }{u_h^2} \right) \right]  
\left[ u_c^2-u_h^2+u_h^2 \cosh \left(  \frac{\sqrt{u_h^2-u_c^2}~x_{43} }{u_h^2} \right) \right]  }{  
\left[ u_c^2-u_h^2+u_h^2 \cosh \left(  \frac{\sqrt{u_h^2-u_c^2}~x_{32} }{u_h^2} \right) \right]  
\left[ u_c^2-u_h^2+u_h^2 \cosh \left(  \frac{\sqrt{u_h^2-u_c^2}~x_{41} }{u_h^2} \right) \right]  }} \ \  \right). \nonumber
\end{align}
To compare with the field theory computations in \cref{sn_dj_ft}, we have to take the limit of small deformation parameter $\mu$, corresponding to large cut-off radius $r_c$ (or small $u_c$) [see \cref{cutoff_rad}]. Further we must consider the high temperature limit $\beta \ll |x_{ij}|$, as the dual cut-off geometry resembles a BTZ black hole only in the high temperature limit. Expanding \cref{EW_dj} for small $u_c$ and $\beta \ll |x_{ij}|$ we arrive at
\begin{align}
\label{EW_dj_lim}
E_W(A:B) & =
\frac{1}{4G_N} \cosh^{-1} \left[  
1 +   2 \frac{
\sinh \left( \frac{x_{21}}{2u_h} \right) \sinh \left( \frac{x_{43}}{2u_h} \right) }{
\sinh \left( \frac{x_{32}}{2u_h} \right) \sinh \left( \frac{x_{41}}{2u_h} \right) }  \right] \\
& - \frac{u_c^2}{16G_N u_h^3} \sqrt{ \frac{
\sinh \left( \frac{x_{21}}{2u_h} \right) \sinh \left( \frac{x_{43}}{2u_h} \right) }{
\sinh \left( \frac{x_{31}}{2u_h} \right) \sinh \left( \frac{x_{42}}{2u_h} \right) }}
\left[  x_{21} \coth \left( \frac{x_{21}}{2u_h} \right) 
+ x_{43} \coth \left( \frac{x_{43}}{2u_h} \right) \right. \nonumber \\
& \left.  -x_{32} \coth \left( \frac{x_{32}}{2u_h} \right) 
- x_{41} \coth \left( \frac{x_{41}}{2u_h} \right)   \right] 
- \frac{u_c^2}{32G_N u_h^2} 
\left(
\sqrt{\frac{
\sinh \left( \frac{x_{31}}{2u_h} \right) \sinh \left( \frac{x_{42}}{2u_h} \right)
}{
\sinh \left( \frac{x_{21}}{2u_h} \right) \sinh \left( \frac{x_{43}}{2u_h} \right)
}} \right. \nonumber \\
& \left. \times \left[
\csch^2 \left( \frac{x_{31}}{2u_h} \right) +\csch^2 \left( \frac{x_{42}}{2u_h} \right) 
-\csch^2 \left( \frac{x_{32}}{2u_h} \right) -\csch^2 \left( \frac{x_{41}}{2u_h} \right) \right] \right. \nonumber \\
& + \sqrt{\frac{
\sinh \left( \frac{x_{21}}{2u_h} \right) \sinh \left( \frac{x_{43}}{2u_h} \right)  }{
\sinh \left( \frac{x_{31}}{2u_h} \right) \sinh \left( \frac{x_{42}}{2u_h} \right)  }} \nonumber \\
& \left. \times \left[
\csch^2 \left( \frac{x_{21}}{2u_h} \right) +\csch^2 \left( \frac{x_{43}}{2u_h} \right)
-\csch^2 \left( \frac{x_{32}}{2u_h} \right) -\csch^2 \left( \frac{x_{41}}{2u_h} \right) \right]
 \right). \nonumber
\end{align}
The first term in \cref{EW_dj_lim} is the EWCS between the two disjoint intervals for the corresponding undeformed \cft{2}. The rest of the terms (proportional to $u_c^2$ and thus to $\mu$) describes the leading order corrections to the EWCS due to the \ttbar{} deformation. The third term becomes negligible (compared to the second term) in the high temperature limit. The change in HEE for two disjoint intervals in proximity due to the \ttbar{} deformation is given by \cite{Jeong:2019ylz}
\begin{align}
\label{chee_dj}
\delta S(A\cup B) = - \frac{\mu c^2 \pi ^4 }{9\beta^3} 
\left[ x_{32} \coth \left( \frac{\pi x_{32} }{\beta} \right)
+ x_{41} \coth \left( \frac{\pi x_{41}}{\beta} \right) \right].
\end{align}
The change in holographic OEE for two disjoint intervals due to the \ttbar{} deformation may now be computed by combining \cref{EW_dj_lim,chee_dj} through \cref{hoee_def}.
Interestingly our holographic result matches exactly with our earlier field theory computation in \cref{coee_dj_x}, in the large central charge limit together with small deformation parameter and high temperature limits, which serves as a strong consistency check for our holographic construction.

\subsubsection{Two adjacent intervals}
\label{sn_adj_ft_hol}

\begin{figure}[h!]
\centering
\includegraphics[scale=0.50]{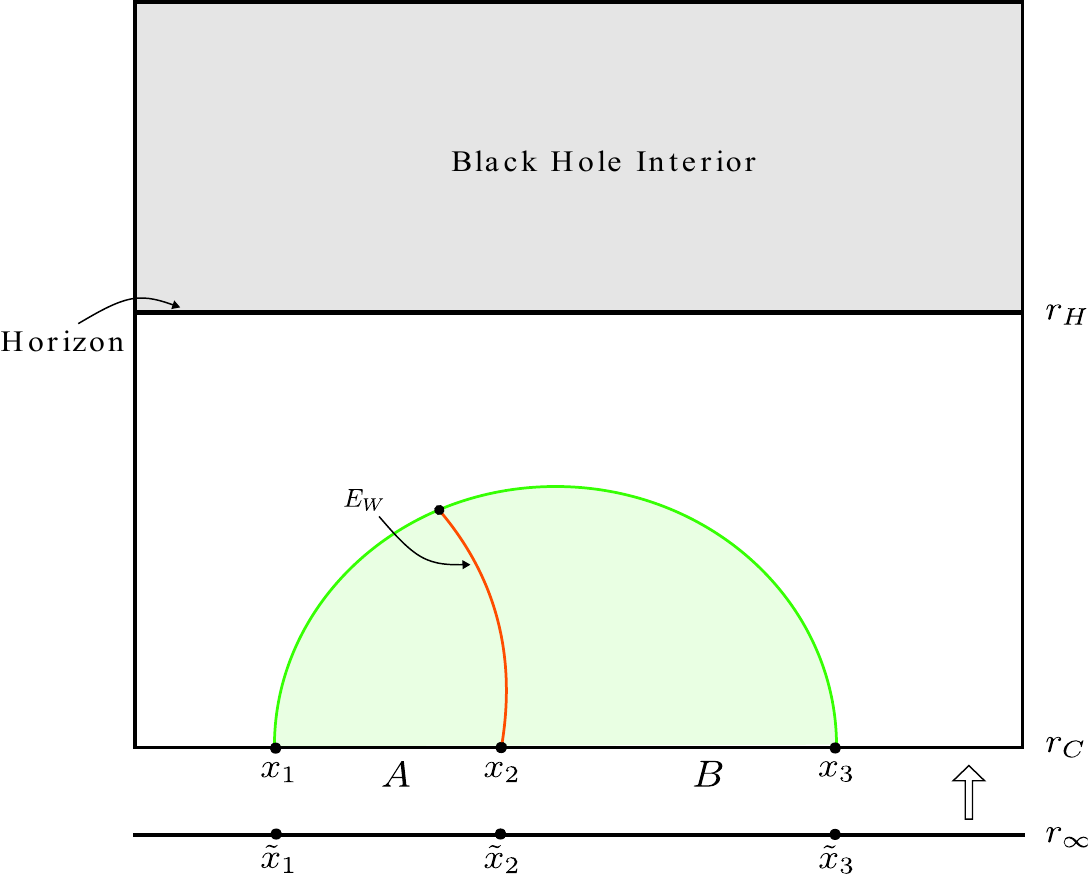}
\caption{EWCS for two adjacent intervals in a $T\bar{T}$ deformed \cft{2}. Figure based on \cite{Basu:2023bov}.}
\label{fg_EW_adj}
\end{figure}

We now consider two adjacent intervals $A=[x_1,x_2]$ and $B=[x_2,x_3]$ with $x_1<x_2<x_3$ as described 
in \cref{sn_adj_ft}. The configuration has been depicted in \cref{fg_EW_adj}. The EWCS for the corresponding bulk points $X(s_1),X(s_2),X(s_3)$ is given by \cite{Kusuki:2019evw}
\begin{align}
\label{EW_adj_def}
E_W=\frac{1}{4 G_N} \cosh ^{-1} \left(\frac{\sqrt{2}}{\sqrt{v}}\right),
\end{align}
where
\begin{align}
v= \frac{\xi_{13}^{-1}}{\xi_{12}^{-1} \xi_{23}^{-1}} \;,\qquad
\xi_{ij}^{-1}=-X(s_i)\cdot X(s_j) \;.
\end{align}
As earlier the three points on the boundary may be expressed in the global coordinates as $X(0,r_c,x_i)$ for $i=1,2,3$. The corresponding EWCS may then be computed from \cref{EW_adj_def} as
\begin{align}
\label{EW_adj}
E_W(A:B) & =
\frac{1}{4 G_N} \log \left[ \frac{4 u_h
\sinh \left( \frac{x_{21}}{2 u_h} \right) \sinh \left( \frac{x_{32}}{2 u_h} \right) }{ u_c
\sinh \left( \frac{x_{31}}{2 u_h} \right) } \right] \\
& - \frac{u_c^2}{16 G_N u_h^3} \left[
x_{21} \coth \left( \frac{x_{21}}{2 u_h} \right) - x_{31} \coth \left( \frac{x_{31}}{2 u_h} \right)
+ x_{32} \coth \left( \frac{x_{32}}{2 u_h} \right) \right] \nonumber \\
& + \frac{u_c^2}{16 G_N u_h^2}\left[ 
\csch^2 \left( \frac{x_{21}}{2 u_h} \right) - \csch^2 \left( \frac{x_{31}}{2 u_h} \right)
+ \csch^2 \left( \frac{x_{32}}{2 u_h} \right) \right]. \nonumber
\end{align}

Similar to the disjoint configuration, the first term in \cref{EW_adj} is the EWCS between the two adjacent intervals for the corresponding undeformed \cft{2}. The rest of the terms (proportional to $u_c^2$ and thus to $\mu$) describes the leading order corrections for the EWCS due to the \ttbar{} deformation. The third term becomes negligible (compared to the second term) in the high temperature limit. The change in HEE for two adjacent intervals due to the \ttbar{} deformation is given by \cite{Jeong:2019ylz}
\begin{align}
\label{chee_adj}
\delta S(A\cup B)= - \left( \frac{\mu c^2 \pi^4 }{9 \beta^3} \right) x_{31}
\coth \left(\frac{\pi  x_{31}}{\beta }\right) .
\end{align}
The change in holographic OEE for two adjacent intervals due to the \ttbar{} deformation may now be obtained from \cref{hoee_def,EW_adj,chee_adj}, and is described by \cref{coee_adj_x}, where as earlier we have used the holographic dictionary. Once again we find exact agreement between our holographic and field theory results (in the large central charge limit, along with small deformation parameter and high temperature limits), which substantiates our holographic construction.

Note that a limiting analysis of the EWCS for two disjoint intervals for the undeformed CFT$_2$ does not lead to the corresponding adjacent result given by the first term in \cref{EW_adj}. This mismatch is not surprising since for the case of disjoint intervals the EWCS is given by a minimal curve between two bulk geodesics whereas for adjacent intervals it is a minimal curve between a bulk geodesic and a boundary point. In this connection, we should not expect the corrections due to the \ttbar{} deformations to have a well defined adjacent limit as well.

\subsubsection{A single interval}
\label{sn_sg_ft_hol}

\begin{figure}[h!]
\centering
\includegraphics[scale=0.50]{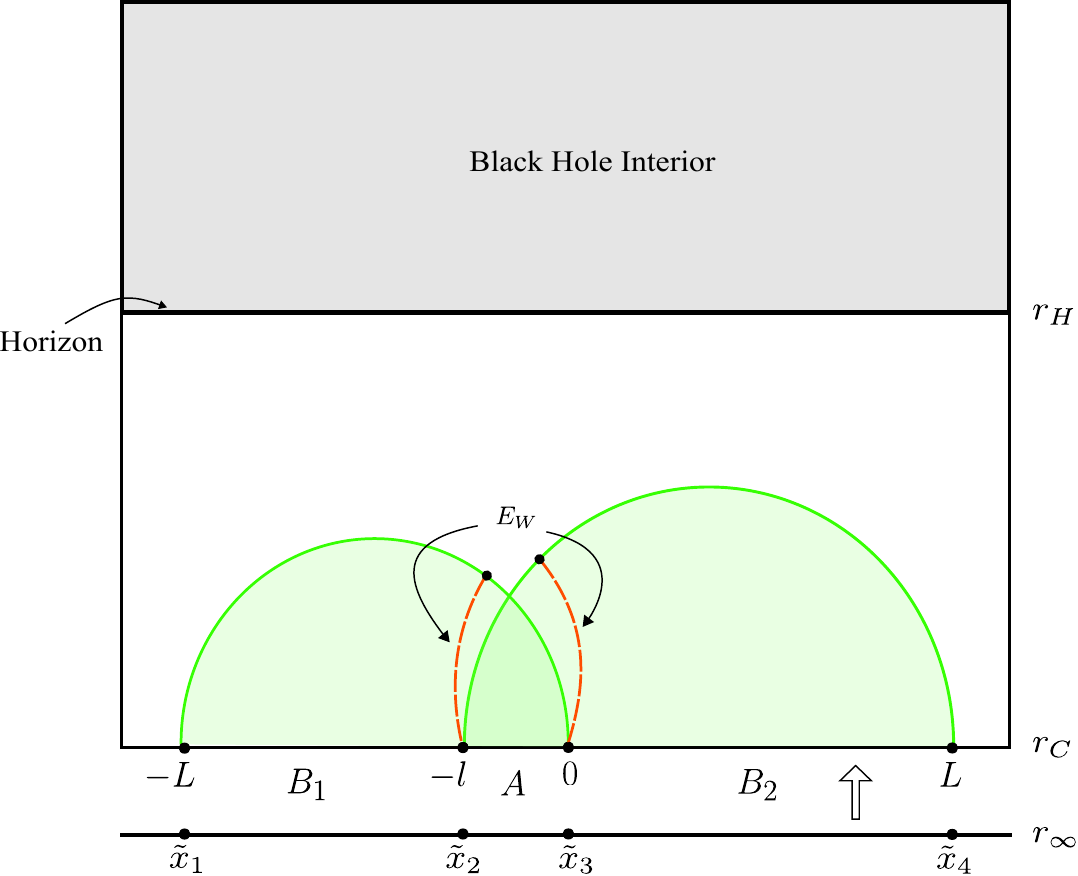}
\caption{EWCS for a single interval in a $T\bar{T}$ deformed \cft{2}. Figure based on \cite{Basu:2023bov}.}
\label{fg_EW_sg}
\end{figure}

Finally we consider the case of a single interval $A=[-\ell,0]$ in a thermal \ttbar{} deformed holographic \cft{2} ($\ell>0$). As described in \cref{sn_sg_ft} this necessitates the introduction of two large but finite auxiliary intervals $B_1=[-L, -\ell]$ and $B_2=[0,L]$ sandwiching the interval $A$ with $B\equiv B_1\cup B_2$ ($L\gg\ell$) \cite{Calabrese:2014yza}. The situation has been outlined in \cref{fg_EW_sg}.

We then compute the holographic OEE for this modified configuration, and finally take the bipartite limit $B\to A^c$ (implemented through $L\to\infty$) to obtain the desired OEE for the original configuration of the single interval $A$. The EWCS between the intervals $A$ and $B=B_1\cup B_2$ may be computed from the following relation 
\cite{KumarBasak:2020eia,Basu:2021awn,Basu:2022nds}
\begin{align}
\label{EW_sg_sum}
\tilde{E}_W(A:B)=E_W(A:B_1)+E_W(A:B_2)\,,
\end{align}
where $\tilde{E}_W(A:B)$ denotes an upper bound on the EWCS between the intervals $A$ and $B$. 
All subsequent computations involving \cref{EW_sg_sum} should be interpreted accordingly. Note that each term on the right hand side of \cref{EW_sg_sum} represents the EWCS of two adjacent intervals which has already been computed in \cref{sn_adj_ft_hol}. The corrections to these terms may thus be read off from \cref{EW_adj} as follows
\begin{align}
\label{EW1_sg}
\delta E_W(A:B_1) = - \frac{u_c^2}{16 G_N u_h^3} \left[
\ell \coth \left( \frac{\ell}{2 u_h} \right)
+ (L-\ell) \coth \left( \frac{L - \ell}{2 u_h} \right)
- L \coth \left( \frac{L}{2 u_h} \right) \right], 
\end{align}
and
\begin{align}
\label{EW2_sg}
\delta E_W(A:B_2) = - \frac{u_c^2}{16 G_N u_h^3} \left[
\ell \coth \left( \frac{\ell}{2 u_h} \right)
+ L \coth \left( \frac{L}{2 u_h} \right)
-(L+\ell) \coth \left( \frac{L+\ell}{2 u_h} \right) \right], 
\end{align}
where we have already taken the limits of small deformation parameter and high temperature. The correction to the HEE for a single interval is given as follows \cite{Jeong:2019ylz}
\begin{align}
\label{chee_sg}
\delta S (A\cup A^c)= - \left( \frac{2 \mu c^2 \pi ^4 L }{9 \beta ^3} \right)
\coth \left(\frac{2 \pi L}{\beta }\right),
\end{align}
where the bipartite limit has already been implemented. The correction to holographic OEE for a single interval due to the \ttbar{} deformation may then be computed from \cref{EW_sg_sum,EW1_sg,EW2_sg,chee_sg} through \cref{hoee_def} on effecting the bipartite limit $L\to\infty$ as follows
\begin{align}
\label{choee_sg}
\delta S_o (A : A^c) = -\frac{\mu c^2 \pi^4 \ell}{9\beta^3} \left[ \coth \left( \frac{\pi \ell}{\beta} \right) - 1 \right]- \lim_{L\to\infty} \left[  \frac{\mu c^2 \pi ^4 L}{9\beta^3} 
\coth \left( \frac{2 \pi L}{\beta} \right) \right] 
,
\end{align}
where we have utilized the holographic dictionary as earlier. Note that on taking the high temperature limit ($\beta\to 0$), \cref{coee_sg} reduces (the second part of the first term becomes negligible as $e^{-\frac{2 \pi \ell}{\beta}}\to 0$) exactly to \cref{choee_sg}. This once again serves as a robust consistency check for our holographic construction. 

We may understand the corrections to the thermal entropy described in \cref{Thermal-entropy} from a holographic viewpoint as well. Recall that the holographic entanglement entropy receives the thermal contribution as the corresponding RT surface wraps the black hole horizon \cite{Hubeny:2007xt}. Under the \ttbar{} deformation, the holographic screen is pushed inside the bulk and the wrapping of the corresponding minimal surface around the black hole horizon is now smaller compared to the undeformed case. As a result, the contribution to the thermal entropy decreases compared to the undeformed case.

\section{\ttbar{} deformed finite size \cft{2} and holography}
\label{sn_fs}

\subsection{OEE in a \ttbar{} deformed finite size \cft{2}}
\label{sn_oee_fs}

In this section we follow a similar prescription as in \cref{sn_oee_ft} to formulate a perturbative expansion for the OEE in a \ttbar{} deformed finite size \cft{2} of length $L$ at zero temperature. For this setup, the corresponding manifold $\mathcal{M}$ describes an infinitely long cylinder of circumference $L$ with the length direction periodically compactified by the relation $x\sim x+L$ \cite{Calabrese:2012nk}. The cylindrical manifold $\mathcal{M}$ for this configuration may be represented by the complex coordinates described in \cref{cyl_coord} with the spatial coordinate $x\in (0,L)$ and the time coordinate $\tau\in (-\infty,\infty)$ \cite{Calabrese:2012nk}. The cylinder $\mathcal{M}$ may be further described on the complex plane $\mathbb{C}$ through the following conformal map \cite{Calabrese:2012nk}
\begin{align}
\label{map_fs}
z=e^{- \frac{2\pi i w}{L}} \;, \qquad\qquad \bar{z}=e^{\frac{2\pi i \bar{w}}{L}} \;,
\end{align}
where $(z, \bar{z})$ are the coordinates on the complex plane. The relations in \cref{stress_trfn,tt_sg_sht} remain valid with $\beta$ effectively replaced by $iL$. With these modifications, the expressions in \cref{ptfn_def,doee_act,doee_exp,coee_def} may now be applied to compute the OEE in a \ttbar{} deformed finite size \cft{2} at zero temperature.

\subsubsection{Two disjoint intervals}
\label{sn_dj_fs}

As earlier we start with the mixed state of two disjoint spatial intervals $A=[x_1,x_2]$ and $B=[x_3,x_4]$ in a \ttbar{} deformed finite size \cft{2} of length $L$ at zero temperature, defined on the cylindrical manifold $\mathcal{M}$ described above ($x_1<x_2<x_3<x_4$). The first order correction in the OEE of two disjoint intervals in a \ttbar{} deformed finite size \cft{2} may be obtained by substituting \cref{tt_dj,ward_hol,tt_dj_2,4pt_fn} along with \cref{map_fs} ($\beta$ replaced by $iL$) into \cref{coee_def} as follows
\begin{align}
\label{coee_dj_fs_int}
\delta S_o(A:B) & = \frac{-\mu c^2 \pi ^4 }{18L^4(z_1-z_3)^2(z_2-z_4)^2(\eta -1)\sqrt{\eta}}    \\
&    \times  \int _{\mathcal{M}} z^2 \left[ \frac{(z_2-z_3)(z_2-z_4)((z-z_1)(z_3-z_4)+(z_1-z_3)(2z-3z_1+z_4) \sqrt{\eta})}{(z-z_1)^2}  \right. \nonumber  \\
&   \left. + \frac{(z_1-z_3)(z_1-z_4)(-((z-z_2)(z_3-z_4))+(2z-3z_2+z_3)(z_2-z_4)\sqrt{\eta})}{(z-z_2)^2} \right. \nonumber  \\ 
&   \left. - \frac{(z_1-z_4)(z_2-z_4)((z_1-z_2)(-z+z_3)+(2z+z_2-3z_3)(z_1-z_3)\sqrt{\eta})}{(z-z_3)^2} \right. \nonumber  \\ 
&   \left. + \frac{(z_1-z_3)(z_3-z_2)((z_1-z_2)(z-z_4)+(2z+z_1-3z4_)(z_2-z_4)\sqrt{\eta})}{(z-z_4)^2}  \right]. \nonumber
\end{align}
We now substitute $ z \to e^{-\frac{2\pi i (x+i\tau)}{L}} $ into \cref{coee_dj_fs_int} and integrate the resulting expression with respect to $x$ to arrive at
\begin{align}
\label{coee_dj_fs_int_t}
\delta S_o(A:B) & = \frac{i\mu c^2\pi ^3}{36L^3 \sqrt{\eta}} \int d\tau \left[ \frac{z_1\sqrt{\eta}}{e^{\frac{2\pi (-ix+\tau)}{L}}-z_1} +  \frac{z_2\sqrt{\eta}}{e^{\frac{2\pi (-ix+\tau)}{L}}-z_2} + \frac{z_3\sqrt{\eta}}{e^{\frac{2\pi (-ix+\tau)}{L}}-z_3}  \right.  \\ 
&  \left. + \frac{z_4\sqrt{\eta}}{e^{\frac{2\pi (-ix+\tau)}{L}}-z_4}
+ \frac{(z_1(z_3-z_4)+(z_1-z_3)(z_1+z_4) \sqrt{\eta})\log [e^{\frac{2\pi (-ix+\tau)}{L}}-z_1]}{(z_1-z_3)(z_1-z_4)} 
\right. \nonumber \\ 
&  \left. + \frac{(z_2(z_4-z_3)+(z_2+z_3)(z_2-z_4) \sqrt{\eta})\log [e^{\frac{2\pi (-ix+\tau)}{L}}-z_2]}
{(z_2-z_3)(z_2-z_4)} \right. \nonumber \\ 
&  \left. + \frac{((z_2-z_1)z_3+(z_1-z_3)(z_2+z_3) \sqrt{\eta})\log [e^{\frac{2\pi (-ix+\tau)}{L}}-z_3]}
{(z_1-z_3)(z_3-z_2)} \right. \nonumber \\ 
&  \left. + \frac{((z_2-z_1)z_4+(z_1+z_4)(z_4-z_2) \sqrt{\eta})\log [e^{\frac{2\pi (-ix+\tau)}{L}}-z_4]}
{(z_1-z_4)(z_4-z_2)}  \right]. \nonumber
\end{align}
We observe that the first four terms on the right hand side of \cref{coee_dj_fs_int_t} readily vanish on inserting the limits of integration $x=0$ and $x=L$. Since we have considered the system on a constant time slice, we may take $\tau_j$ ($j=1,2,3,4$) to be zero for all boundary points, and the contributions of the logarithmic functions become zero identically. Thus it is observed that the resultant integrand for the $\tau$ integration in \cref{coee_dj_fs_int_t} vanishes leading to no non-trivial first order correction to the OEE. This is in conformity with the vanishing entanglement entropy for a finite sized \ttbar{} deformed CFT$_2$ \cite{Chen:2018eqk}.

\subsubsection{Two adjacent intervals}
\label{sn_adj_fs}

We now focus on the bipartite mixed state of two adjacent intervals $A=[x_1,x_2]$ and $B=[x_2,x_3]$ in a \ttbar{} deformed finite size \cft{2} of length $L$ at zero temperature, defined on the cylindrical manifold $\mathcal{M}$ described by \cref{cyl_coord,map_fs} ($x_1<x_2<x_3$). For this case, \cref{coee_def,tt_adj,tt_adj_2,3pt_fn} may still be employed along with the relation described in \cref{map_fs}, effectively replacing $\beta$ by $iL$. The first order correction in OEE due to $\mu$ for two adjacent intervals is then given by
\begin{align}
\label{coee_adj_fs_int}
\delta S_o(A:B) & = -\frac{\mu c^2\pi^4}{18L^4} \int_{\mathcal{M}} \frac{z^2}{(z-z_1)^2(z-z_2)^2(z-z_3)^2} \\
& \times \left[ z_2^2z_3^2-z_1z_2z_3(z_2+z_3)+z_1^2(z_2^2-z_2z_3+z_3^2) \right. \nonumber \\ 
& \left. +z^2(z_1^2+z_2^2-z_2z_3+z_3^2-z_1(z_2+z_3))-z(z_1^2(z_2+z_3) \right. \nonumber \\
& \left. +z_2z_3(z_2+z_3)+z_1(z_2^2-6z_2z_3+z_3^2)) \right]. \nonumber
\end{align}
Next we replace $ z \to e^{-\frac{2\pi i (x+i\tau)}{L}} $ into \cref{coee_adj_fs_int} and subsequently integrate with respect to $x$ to obtain
\begin{align}
\label{coee_adj_fs_int_t}
\delta S_o(A:B) & = \frac{i\mu c^2\pi^3}{36L^3} \int d\tau  \left[ \frac{z_1}{e^{\frac{2\pi (-ix+\tau)}{L}}-z_1}+\frac{z_2}{e^{\frac{2\pi (-ix+\tau)}{L}}-z_2}+\frac{z_3}{e^{\frac{2\pi (-ix+\tau)}{L}}-z_3} \right. \\
& \left. +\frac{(z_1^2-z_2z_3)\log [e^{\frac{2\pi (-ix+\tau)}{L}}-z_1]}{(z_1-z_2)(z_1-z_3)} +\frac{(z_2^2-z_1z_3)\log [e^{\frac{2\pi (-ix+\tau)}{L}}-z_2]}{(z_2-z_1)(z_2-z_3)} \right. \nonumber \\
& \left.  +\frac{(z_3^2-z_2z_1)\log [e^{\frac{2\pi (-ix+\tau)}{L}}-z_3]}{(z_1-z_3)(z_2-z_3)}  \right]. \nonumber
\end{align}
Similar to the disjoint case, the first three terms on the right hand side of \cref{coee_adj_fs_int_t} readily vanish when the limits of integration $x=0$ and $x=L$ are inserted. As earlier, for a constant time slice $\tau_j=0$ ($j=1,2,3$), the
logarithmic functions also contribute nothing to the definite integral. The resulting integrand for the $\tau$ integration in \cref{coee_adj_fs_int_t} thus vanishes. Hence the corresponding first order correction in the OEE of two adjacent intervals turns out to be zero.

\subsubsection{A single interval}
\label{sn_sg_fs}

Finally we turn our attention to the bipartite mixed state configuration of a single interval $A=[x_1,x_2]$ in a \ttbar{} deformed finite size \cft{2} of length $L$ at zero temperature, defined on the cylindrical manifold $\mathcal{M}$ given in \cref{cyl_coord,map_fs} ($x_1<x_2$). The construction of the relevant partially transposed reduced density matrix for this configuration is described in \cite{Calabrese:2012nk}. Once again we may utilize \cref{tt_sg,tt_sg_2} with only two points $z_1$ and $z_2$, subject to \cref{map_fs} (with the effect of $iL$ replacing $\beta$), and a two point twist correlator as mentioned below in \cref{2pt_fn}. We have expressed the modified version of \cref{tt_sg_2} as applicable for the system under consideration for convenience of the reader as follows
\begin{align}
\label{tt_sg_fs}
\int_{\mathcal{M}_{n_o}} \Braket{T\bar{T}}_{\mathcal{M}_{n_o}}
& = \frac{1}{n_o} \int_{\mathcal{M}} \frac{1}{\Braket{ \sigma^{2}_{n_o}(z_{1}, \bar{z}_{1})\bar{\sigma}^{2}_{n_o}(z_{2}, \bar{z}_{2}) } } \\
& \times \left[ \frac{\pi^2 c\,n_o}{6 L^2} - \left(\frac{2 \pi z}{L} \right)^2 \sum_{j=1}^{2} \left(\frac{h_j}{(z-z_j)^2}+\frac{1}{(z-z_j)}\partial_{z_j}\right) \right] \nonumber \\
& \times \left[ \frac{\pi^2 c\,n_o}{6 L^2} - \left(\frac{2 \pi \bar{z}}{L} \right)^2 \sum_{k=1}^{2} \left(\frac{\bar{h}_k}{(\bar{z}-\bar{z}_k)^2}+\frac{1}{(\bar{z}-\bar{z}_k)}\partial_{\bar{z}_k}\right) \right] \nonumber \\
& \times\Braket{ \sigma^{2}_{n_o}(z_{1}, \bar{z}_{1})\bar{\sigma}^{2}_{n_o}(z_{2}, \bar{z}_{2}) }_{\mathcal{C}} \,, \nonumber
\end{align}
where $h_1=h_2=h^{(2)}_{n_o}$ with $\bar{h}_{i}=h_{i} \; (i=1,2)$ [see \cref{sigma_dim,sigma_sq_dim}]. The corresponding two point twist correlator for this configuration is given by \cite{Calabrese:2012nk}
\begin{align}
\label{2pt_fn}
\Braket{\sigma^{2}_{n_o}(z_{1}, \bar{z}_{1})\bar{\sigma}^{2}_{n_o}(z_{2}, \bar{z}_{2})}
=  \frac{\mathcal{C}_{12}}{\left| z_1-z_2 \right|^{2h_{n_o}}} \;,
\end{align}
where $\mathcal{C}_{12}$ is the relevant normalization constant. Following a similar procedure like the earlier cases, the first order correction for the OEE of this setup may be given as follows
\begin{align}
\label{coee_sg_fs_int}
\delta S_o(A:B)= -\frac{\mu c^2\pi^4}{18L^4} (z_1-z_2)^2 \int_{\mathcal{M}} \frac{z^2}{(z-z_1)^2(z-z_2)^2} \;.
\end{align}
We then obtain the following expression by substituting $ z \to e^{-\frac{2\pi i (x+i\tau)}{L}} $ into \cref{coee_sg_fs_int} and integrating with respect to $x$
\begin{align}
\label{coee_sg_fs_int_t}
\delta S_o(A:B) &= \frac{i \mu c^2\pi^3}{36L^3} \int d\tau \left[ \frac{z_1}{ e^{\frac{2\pi (-ix+\tau)}{L}}-z_1}+\frac{z_2}{ e^{\frac{2\pi (-ix+\tau)}{L}}-z_2} \right. \\
& \left. +\frac{z_1+z_2}{z_1-z_2} \left( \log \left[ e^{\frac{2\pi (-ix+\tau)}{L}}-z_1 \right]-\log \left[ e^{\frac{2\pi (-ix+\tau)}{L}}-z_2 \right] \right)   \right]. \nonumber
\end{align}
Like the previous cases, we observe that the first two terms in \cref{coee_sg_fs_int_t} vanish on implementation of the limits of integration $x=0$ and $x=L$. As the system under consideration is on a constant time slice $\tau_j=0$ ($j=1,2$), once again the terms containing the logarithmic functions also vanish. Again the resulting integrand for the $\tau$ integration in \cref{coee_sg_fs_int_t} vanishes, indicating the vanishing of the first order corrections of the OEE as earlier.

\subsection{Holographic OEE in a \ttbar{} deformed finite size \cft{2}}
\label{sn_hoee_fs}

The bulk dual of a \ttbar{} deformed finite size \cft{2} of length $L$ at zero temperature is represented by a finite cut-off \ads{3} geometry expressed in global coordinates as follows \cite{Ryu:2006bv,Ryu:2006ef}
\begin{align}
\label{global_metric}
\mathrm{d}s^2=R^2 \left( -\cosh^2\rho\, \mathrm{d}\tau^2 +\sinh^2 \rho\, \mathrm{d}\phi^2 + \mathrm{d}\rho^2  \right),
\end{align}
where $\phi=2\pi x/L$. As earlier we embed this \ads{3} geometry in $\mathbb{R}^{2,2}$ as follows \cite{Kusuki:2019evw}
\begin{align}
\label{embed_fs}
\mathrm{d}s^2=\eta_{AB} \mathrm{d}X^A \mathrm{d}X^B
=-\mathrm{d}X^2_0-\mathrm{d}X^2_1+\mathrm{d}X^2_2+\mathrm{d}X^2_3\:,
\qquad X^2=-1\,.
\end{align}
The metric in \cref{global_metric} may be expressed in terms of the embedding coordinates introduced in \cref{embed_fs} as follows
\begin{align}
\label{global_coord_fs}
X_0(\tau,\phi,\rho) &= R \cosh \rho \sin \tau, & X_1(\tau,\phi,\rho) &= R \cosh \rho \cos \tau, \\
X_2(\tau,\phi,\rho) &= R \sinh \rho \cos \phi, & X_3(\tau,\phi,\rho) &= R \sinh \rho \sin \phi. \nonumber
\end{align}
The finite cut-off of the \ads{3} geometry is located at $\rho=\rho_c$, where
\begin{align}
\label{rho_c_mu}
\cosh \rho_c = \sqrt{\frac{3L^2}{2 \mu c \pi^3}} \;.
\end{align}
With the UV cut-off of the field theory given by $\epsilon = \sqrt{\mu c \pi / 6}$ [see \cref{cutoff_rad}], the relation in \cref{rho_c_mu} may be rewritten as
\begin{align}
\label{rho_c_ep}
\cosh \rho_c=\frac{L}{2 \pi \epsilon} \; .
\end{align}

\subsubsection{Two disjoint intervals}
\label{sn_dj_fs_hol}

We begin with two disjoint spatial intervals $A=[x_1,x_2]$ and $B=[x_3,x_4]$ on a cylindrical manifold $\mathcal{M}$ as detailed in \cref{sn_dj_fs} ($x_1<x_2<x_3<x_4$). Note that the EWCS involving arbitrary bulk points $X(s_1),X(s_2),X(s_3),X(s_4)$ for a \ttbar{} deformed finite size \cft{2} is described by \cite{Kusuki:2019evw}
\begin{align}
\label{EW_dj_fs_def}
E_W
=\frac{1}{4G_N} \cosh ^{-1} \left(  \frac{1+\sqrt{u}}{\sqrt{v}}  \right),
\end{align}
where
\begin{align}
u=\dfrac{\xi^{-1}_{12}\xi^{-1}_{34}}{\xi^{-1}_{13}\xi^{-1}_{24}}\;,\qquad v=\dfrac{\xi^{-1}_{14}\xi^{-1}_{23}}{\xi^{-1}_{13}\xi^{-1}_{24}}\;,\qquad
\xi^{-1}_{ij}=-X(s_i)\cdot X(s_j)\;.
\end{align}
The end points of the two disjoint intervals under consideration on the boundary may be represented by the embedding coordinates as
$X(0,\phi_i,\rho_c)$ for $i=1,2,3,4$, where $\phi_1<\phi_2<\phi_3<\phi_4$ (Note that $\phi_i=2\pi x_i/L$). The corresponding EWCS may then be computed from \cref{EW_dj_fs_def} as
\begin{align}
\label{EW_dj_fs}
E_W(A:B) &= \frac{1}{4G_N} \cosh^{-1} \left (  \sqrt{\frac{\left[ 1+
\sin^2\left( \frac{\pi x_{31}}{L}\right) \sinh^2\rho_c \right] \left[ 1+
\sin^2\left( \frac{\pi x_{42}}{L}\right) \sinh^2\rho_c \right]}{\left[ 1+
\sin^2\left( \frac{\pi x_{32}}{L}\right) \sinh^2\rho_c \right] \left[ 1+
\sin^2\left( \frac{\pi x_{41}}{L}\right) \sinh^2\rho_c \right]}} \right. \\
& \left. + \sqrt{\frac{\left[ 1+
\sin^2\left( \frac{\pi x_{21}}{L}\right) \sinh^2\rho_c \right] \left[ 1+
\sin^2\left( \frac{\pi x_{43}}{L}\right) \sinh^2\rho_c \right]}{\left[ 1+
\sin^2\left( \frac{\pi x_{32}}{L}\right) \sinh^2\rho_c \right] \left[ 1+
\sin^2\left( \frac{\pi x_{41}}{L}\right) \sinh^2\rho_c \right]}} \;\; \right). \nonumber
\end{align}
To extract the desired first order corrections, we now expand \cref{EW_dj_fs} in small $(1/\cosh \rho_c)$ as follows
\begin{align}
\label{EW_dj_fs_lim}
&E_W(A:B)=
\frac{1}{4G_N}  \cosh^{-1} \left[ 1 + 2\frac{ 
\sin \left( \frac{\pi x_{21}}{L}\right) \sin \left( \frac{\pi x_{43}}{L} \right) }{
\sin \left( \frac{\pi x_{32}}{L}\right) \sin \left( \frac{\pi x_{41}}{L} \right)} \right] +\mathcal{O} \left[\epsilon^2 \right],
\end{align}
where we have utilized \cref{rho_c_ep} to substitute $\epsilon$. The first term in \cref{EW_dj_fs_lim} is the EWCS between the two disjoint intervals for the corresponding undeformed \cft{2}. The rest of the terms characterizing the corrections for the EWCS due to the \ttbar{} deformation are second order and higher in $\epsilon$ and thus negligible. The corresponding leading order corrections for the HEE due to the \ttbar{} deformation has been shown to be zero \cite{Chen:2018eqk}. Thus the leading order corrections to the holographic OEE of two disjoint intervals in a \ttbar{} deformed finite size \cft{2} is zero, which is in complete agreement with our corresponding field theory computations in the large central charge limit described in \cref{sn_dj_fs}.

The vanishing of the EWCS as well as the entanglement entropy may be attributed to the fact that in \ttbar{} deformed finite sized CFT$_2$s, the lengths of the intervals do not depend on the cutoff radius in \cref{rho_c_mu}. In contrast, for thermal CFT$_2$s the lengths of the intervals depend non-trivially (cf. \cref{cft_metric}) on the cutoff radius $r_c$ as long as $r_h\neq 0$ (or, $1/\beta\neq 0$) \cite{Chen:2018eqk}. We will discuss this issue further in \cref{sn_sum}.

\subsubsection{Two adjacent intervals}
\label{sn_adj_fs_hol}

We now turn our attention to the case of two adjacent intervals $A=[x_1,x_2]$ and $B=[x_2,x_3]$ ($x_1<x_2<x_3$) as described in \cref{sn_adj_fs}.
The bulk description of the end points of the intervals $A$ and $B$ for a \ttbar{} deformed finite size \cft{2} is given by
$X(0,\phi_i,\rho_c)$ for $i=1,2,3$, where $\phi_1<\phi_2<\phi_3$ ($\phi_i=2\pi x_i/L$). The EWCS for this configuration is described as follows \cite{Kusuki:2019evw}
\begin{align}
\label{EW_adj_fs_def}
E_W=\frac{1}{4 G_N} \cosh^{-1} \left(\frac{\sqrt{2}}{\sqrt{v}}\right),
\end{align}
where
\begin{align}
v= \frac{\xi_{13}^{-1} }{\xi_{12}^{-1} \xi_{23}^{-1}}\;,\qquad
\xi_{ij}^{-1}=-X(s_i)\cdot X(s_j)\;.
\end{align}
We now utilize \cref{EW_adj_fs_def} to explicitly compute the EWCS as follows
\begin{align}
\label{EW_adj_fs}
& E_W (A:B) \\
& = \frac{1}{4G_N}\cosh ^{-1}\left(
\sqrt{ \frac{ 2 \left[\cosh[2](\rho_c) - \cos (\frac{2\pi x_{21}}{L}) \sinh[2](\rho_c)\right]
\left[\cosh[2](\rho_c) - \cos (\frac{2\pi x_{32}}{L}) \sinh[2](\rho_c)\right] }{ 
\cosh[2](\rho_c) - \cos (\frac{2\pi x_{31}}{L}) \sinh[2](\rho_c) } } \right). \nonumber
\end{align}
We are now in a position to extract the leading order corrections to the EWCS from \cref{EW_adj_fs} by expanding in small $(1/\cosh \rho_c)$ as follows
\begin{align}
\label{EW_adj_fs_lim}
& E_W(A:B) = \frac{1}{4 G_N}\log \left[ \left( \frac{2L}{\pi \epsilon} \right) \frac{ \sin \left(\frac{\pi x_{21}}{L}\right) \sin \left(\frac{\pi x_{32}}{L}\right)}{\sin \left(\frac{\pi x_{31}}{L}\right)}\right]  +\mathcal{O} \left[\epsilon^2 \right],
\end{align}
where we have already substituted the relation in \cref{rho_c_ep}. As earlier the first term on the right hand side of \cref{EW_adj_fs_lim} describes the EWCS between the two adjacent intervals for the corresponding undeformed \cft{2}. Again the \ttbar{} correction terms are second order and higher in $\epsilon$ and negligible. The leading order corrections of the HEE for this configuration due to the \ttbar{} deformation has been demonstrated to be vanishing \cite{Chen:2018eqk}. Hence the leading order corrections to the holographic OEE for this case vanishes, which once again is in conformity with our field theory results in the large central charge limit described in \cref{sn_adj_fs}.

\subsubsection{A single interval}
\label{sn_sg_fs_hol}

The bulk representation of the end points of a single interval of length $\ell$ may be given by $X(0,0,\rho_c)$ and $X(0,\delta\phi,\rho_c)$, where $\delta\phi=\frac{2\pi \ell}{L}$. The EWCS for the given configuration (same as the HEE for a single interval) may be computed as
\begin{align}
\label{EW_sg_fs}
E_W(A:A^c)=\frac{1}{4G_N } \cosh ^{-1}\left[ 1 + 2 \sinh ^2\left(\rho _c\right) \sin ^2\left(\frac{\pi \ell}{L}\right)\right].
\end{align}
Once again \cref{EW_sg_fs} may be expanded for small $(1/\cosh \rho_c)$ to obtain the following expression for the EWCS
\begin{align}
\label{EW_sg_fs_lim}
E_W(A:A^c)=
\frac{1}{2 G_N}\log \left[ \frac{L}{\pi \epsilon} \sin \left(\frac{\pi \ell}{L}\right)\right]
+\mathcal{O} \left[\epsilon^2 \right],
\end{align}
where we have used \cref{rho_c_ep} to replace $\cosh \rho_c$. Once again the first term of \cref{EW_sg_fs_lim} represents the EWCS of a single interval for the corresponding undeformed \cft{2}, while we have neglected the second and higher order correction terms in $\epsilon$. The corresponding corrections for the HEE of a single interval has been shown to be zero \cite{Chen:2018eqk}. Thus the leading order corrections to the holographic OEE for a single interval vanishes, demonstrating agreement with our field theory calculations in the large central charge limit detailed in \cref{sn_sg_fs}.


\section{Summary and discussions}
\label{sn_sum}

To summarize we have computed the OEE for different bipartite mixed state configurations in a \ttbar{} deformed finite temperature \cft{2} with a small deformation parameter $\mu$. In this context we have developed a perturbative construction to compute the first order correction to the OEE for small deformation parameter through a suitable replica technique. This incorporates definite integrals of the expectation value of the \ttbar{} operator over an $n_o$ sheeted replica manifold. We have been able to express these expectation values in terms of appropriate twist field correlators for the configurations under consideration. Utilizing our perturbative construction we have subsequently computed the OEE for the mixed state configurations described by two disjoint intervals, two adjacent intervals, and a single interval in a \ttbar{} deformed thermal \cft{2}.

Following the above we have computed the corresponding EWCS in the dual bulk finite cut-off BTZ black hole geometry for the above configurations utilizing an embedding coordinate technique in the literature. Interestingly it was possible to demonstrate that the first order correction to the sum of the EWCS and the corresponding HEE matched exactly with the first order correction to the \cft{2} replica technique results for the OEE in the large central charge and high temperature limit. This extends the holographic duality for the OEE proposed in the literature to \ttbar{} deformed thermal \cft{2}s.

Finally we have extended our perturbative construction to \ttbar{} deformed finite size \cft{2}s at zero temperature. We have computed the first order corrections to the OEE for the configurations mentioned earlier in such \cft{2}s in the large central charge limit. In all the cases we have been able to show that the leading order corrections vanish in the appropriate limits. Quite interestingly it was possible to demonstrate that the first order corrections to the corresponding bulk EWCS in the dual cut-off BTZ geometry were also identically zero in a further validation of the extension of the holographic duality for the OEE in the literature to \ttbar{} deformed finite size \cft{2}s at zero temperature. 

\begin{figure}[h!]
	\centering
	\includegraphics[scale=0.75]{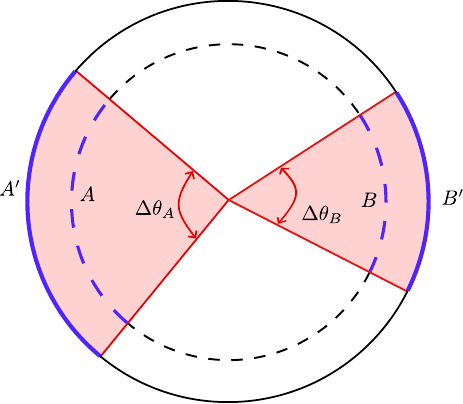}
	\caption{Schematics of two disjoint intervals placed along the compactified direction in a holographic \ttbar{} deformed CFT$_2$. The undashed (dashed) circle denotes the location of the holographic screens before (after) the \ttbar{} deformation.}
	\label{fg_EW_L}
\end{figure}

There are several recurring features of our results. Note that when the intervals are located along a compactified direction, there are no \ttbar{} corrections as the angular separations of the subsystems are not affected by pushing the holographic screen inside the bulk \cite{Jiang:2023ffu}, as depicted in \cref{fg_EW_L}. On the other hand, when this direction is non-compact, the spatial extents of the subsystems become dependent on the finite cutoff radius as depicted in \cref{fg_EW_adj,fg_EW_dj,fg_EW_sg}, and hence there will be appropriate \ttbar{} corrections \cite{Jiang:2023ffu}. For thermal {\cft{2}}s, the time direction is compactified, but the intervals are spatial and hence situated along the spatial direction. Thus for thermal {\cft{2}}s our results indicated corrections due to the \ttbar{} deformations. For finite size {\cft{2}}s, the space direction is compactified, and the corresponding corrections vanish. 

It will be instructive to develop similar constructions for other entanglement measures such as entanglement of purification, balanced partial entanglement, reflected entropy etc. for \ttbar{} deformed \cft{2}s. Also a covariant framework for holographic entanglement in these theories along the lines of the HRT construction is an important open issue. Furthermore, it will be interesting to extend our analysis to \ttbar{} deformations of thermal CFT$_2$ with conserved charges. These constitute exciting open problems for the future.

\section*{Acknowledgments}

We would like to thank Lavish, Mir Afrasiar and Himanshu Chourasiya for valuable discussions. The work of Gautam Sengupta is supported in part by the Dr.\@ Jag Mohan Garg Chair Professor position at the Indian Institute of Technology, Kanpur. The work of Saikat Biswas is supported by the Council of Scientific and Industrial Research (CSIR) of India under Grant No. 09/0092(12686)/2021-EMR-I.

\appendix
\section{The integrals for thermal \cft{2}s}
\label{app_int_ft}

The detailed derivation of the integrals appearing in \cref{coee_dj_int,coee_adj_int,coee_sg_int} has been provided in this appendix. Note that the corresponding domain of integration for all the configurations is the cylindrical manifold $\mathcal{M}$ characterized by the complex coordinates $(w, \bar{w})$ [see \cref{cyl_coord,map}].

\subsection{Two disjoint intervals}
\label{app_dj}

The holomorphic part of the integral in \cref{coee_dj_int} may be written as
\begin{align}
& - \frac{\mu c^2 \pi^4 \sqrt{\eta}}{18 \beta^4 z_{21} z_{32} z_{41} z_{43} }\int_{\mathcal{M}} d^2w \, (z^2) \Bigg[
\frac{z_{32} z_{42} [z_{31} (2z-3z_1+z_4)\sqrt{\eta}+z_{43} (z-z_1)]}{(z-z_1)^2} \\
& + \frac{z_{31} z_{41} [z_{42} (2z-3z_2+z_3)\sqrt{\eta}-z_{43} (z-z_2)]}{(z-z_2)^2}
-\frac{z_{42} z_{41} [z_{31}(2z+z_2-3z_3) \sqrt{\eta}-z_{21}(z-z_3)]}{(z-z_3)^2}  \nonumber \\
& -\frac{z_{31} z_{32} [z_{42} (2z+z_1-3z_4)\sqrt{\eta}+z_{21} (z-z_4)]}{(z-z_4)^2} \Bigg] \nonumber \\
& = -\frac{\mu c^2 \pi ^4  \sqrt{\eta}}{18 \beta^4 z_{21} z_{32} z_{41} z_{43} }\int_0 ^{\infty} dx \int_0 ^{\beta} d\tau \, e^{\frac{4 \pi (x+i\tau)}{\beta}} \\
& \times \left[  \frac{z_{32} z_{42} [z_{31} (2e^{\frac{2\pi(x+i \tau)}{\beta}}-3z_1+z_4)\sqrt{\eta}+z_{43} (e^{\frac{2\pi(x+i \tau)}{\beta}}-z_1)]}{(e^{\frac{2\pi(x+i \tau)}{\beta}}-z_1)^2}	\right. \nonumber \\
& + \frac{z_{31} z_{41} [z_{42} (2e^{\frac{2\pi(x+i \tau)}{\beta}}-3z_2+z_3)\sqrt{\eta}-z_{43} (e^{\frac{2\pi(x+i \tau)}{\beta}}-z_2)]}{(e^{\frac{2\pi(x+i \tau)}{\beta}}-z_2)^2} \nonumber \\
& -\frac{z_{42} z_{41} [z_{31}(2e^{\frac{2\pi(x+i \tau)}{\beta}}+z_2-3z_3) \sqrt{\eta}-z_{21}(e^{\frac{2\pi(x+i \tau)}{\beta}}-z_3)]}{(e^{\frac{2\pi(x+i \tau)}{\beta}}-z_3)^2} \nonumber \\
& \left. -\frac{z_{31} z_{32} [z_{42} (2e^{\frac{2\pi(x+i \tau)}{\beta}}+z_1-3z_4)\sqrt{\eta}+z_{21} (e^{\frac{2\pi(x+i \tau)}{\beta}}-z_4)]}{(e^{\frac{2\pi(x+i \tau)}{\beta}}-z_4)^2} \right]. \nonumber
\end{align}
The primitive function on indefinite integration with respect to $\tau$ turns out to be
\begin{align}
& -\frac{i \mu c^2 \pi^3 }{36 \beta ^3 \sqrt{\eta }}  \left[ \frac{\left(\sqrt{\eta } z_1^2+\left(\sqrt{\eta }-1\right) z_1 (z_{43})-\sqrt{\eta } z_3 z_4\right) \log \left(-z_1+e^{\frac{2 \pi  (x+i \tau )}{\beta }}\right)}{z_{31} z_{41}} \right. \\
& +\frac{\left(\sqrt{\eta } z_2^2+\left(\sqrt{\eta }-1\right) z_2 z_{34}-\sqrt{\eta } z_3 z_4\right) \log \left(-z_2+e^{\frac{2 \pi  (x+i \tau )}{\beta }}\right)}{z_{32} z_{42}} \nonumber \\
& - \frac{\left(\sqrt{\eta } z_1 z_2+\left(\sqrt{\eta }-1\right) z_1 z_3+z_3 \left(-\sqrt{\eta } z_2+z_2-\sqrt{\eta } z_3\right)\right) \log \left(-z_3+e^{\frac{2 \pi  (x+i \tau )}{\beta }}\right)}{z_{31} z_{32}} \nonumber \\
& \left. +\frac{\left(z_4 \left(-\sqrt{\eta } z_2+z_2+\sqrt{\eta } z_4\right)-z_1 \left(\sqrt{\eta } z_2-\sqrt{\eta } z_4+z_4\right)\right) \log \left(-z_4+e^{\frac{2 \pi  (x+i \tau )}{\beta }}\right)}{z_{41} z_{42}} \right] . \nonumber
\end{align}
Due to the presence of branch points, the logarithmic functions necessitate careful treatment while implementing the limits of integration $\tau=0$ and $\tau=\beta$. The following relation outlines the contribution due to a branch point at $z=z_j$ \cite{Chen:2018eqk,Jeong:2019ylz}
\begin{align}
\label{log_identity}
\log\left(e^{\frac{2\pi(x+i \tau)}{\beta}}-z_j\right) \bigg|_{\tau=0}^{\tau=\beta} = \left\{ \begin{array}{ll}
2 \pi i, & \textrm{$\quad \text{for}\; e^{\frac{2\pi x}{\beta}} > z_j \Leftrightarrow x > \frac{\beta}{2\pi} \log z_j$\,,}\\
0, & \textrm{$\quad \text{otherwise.}$}
\end{array} \right.
\end{align}
The branch cuts of the logarithmic functions change the limits of the $x$ integrals as follows
\begin{align}
\int_{-\infty}^{\infty} dx\to \int_{\frac{\beta}{2\pi} \log z_j}^{\infty} dx,
\qquad \text{for} \; j=1,2,3,4. \nonumber
\end{align}
We are now in a position to integrate over $x$ and utilize the prescription described above to implement the limits of integration to arrive at
\begin{align}
& \frac{\mu c^2 \pi^3 }{36 \beta^2} \left(  \frac{\left( z_1 \left( 1+ \sqrt{ \frac{z_{42} z_{43}}{z_{21} z_{31}}} +z_4 \right) \right)}{z_{41}} \log \left[ \frac{z_1}{z_2} \right] +\frac{\left(-2+ \sqrt{\frac{z_{21} z_{43}}{z_{31} z_{42}}} \right) (z_1 z_2-z_3 z_4)}{z_{32} z_{41}} \log \left[ \frac{z_2}{z_3} \right] \right. \\
& \left. + \frac{\left( z_1 +\left( 1+ \sqrt{ \frac{z_{12} z_{31}}{z_{42} z_{43}}} z_4 \right) \right)}{z_{41}} \log \left[ \frac{z_3}{z_4} \right] \right). \nonumber
\end{align}
The anti holomorphic part of the integral in \cref{coee_dj_int} follows a similar analysis and produces the same result as the holomorphic part.

\subsection{Two adjacent intervals}
\label{app_adj}

The holomorphic part of the integral in \cref{coee_adj_int} may be written as
\begin{align}
& \int_{\mathcal{M}} ~ z^2 \left[ \frac{1}{(z-z_1)^2}+\frac{1}{(z-z_2)^2}
+\frac{1}{(z-z_3)^2}+\frac{(-3 z+z_1+z_2+z_3) }{(z-z_1) (z-z_2) (z-z_3)}\right] \\
& =\int_{-\infty}^{\infty}dx\int_0^{\beta}d\tau\; e^{\frac{4 \pi  (x+i \tau )}{\beta }}
\left[ \frac{1}{\left(e^{\frac{2 \pi  (x+i \tau )}{\beta }}-z_1\right)^2}+\frac{1}{\left(e^{\frac{2 \pi  (x+i \tau )}{\beta }}-z_2\right)^2}+\frac{1}{\left(e^{\frac{2 \pi  (x+i \tau )}{\beta }}-z_3\right)^2} \right. \nonumber \\
& \left. +\frac{z_1+z_2+z_3-3\, e^{\frac{2 \pi  (x+i \tau )}{\beta }}}{\left(e^{\frac{2 \pi  (x+i \tau )}{\beta }}-z_1\right) \left(e^{\frac{2 \pi  (x+i \tau )}{\beta }}-z_2\right) 
\left(e^{\frac{2 \pi  (x+i \tau )}{\beta }}-z_3\right)}\right] . \nonumber
\end{align}
We proceed in a similar manner to the disjoint configuration as described in \cref{app_dj}. The indefinite integration with respect to $\tau$ leads to the following primitive function
\begin{align}
& \frac{z_1}{e^{\frac{2 \pi  (x+i \tau )}{\beta }}-z_1}+\frac{z_2}{e^{\frac{2 \pi  (x+i \tau )}{\beta }}-z_2}+\frac{z_3}{e^{\frac{2 \pi  (x+i \tau )}{\beta }}-z_3}+\frac{\left(z_1^2-z_2 z_3\right)}{\left(z_1-z_2\right) \left(z_1-z_3\right)}\log \left(e^{\frac{2 \pi  (x+i \tau )}{\beta }}-z_1\right) \\
& +\frac{\left(z_1 z_3-z_2^2\right)}{\left(z_1-z_2\right)\left(z_2-z_3\right)}\log \left(e^{\frac{2 \pi  (x+i \tau )}{\beta }}-z_2\right)+\frac{\left(z_3^2-z_1 z_2\right) }{\left(z_1-z_3\right)\left(z_2-z_3\right)} \log \left(e^{\frac{2 \pi  (x+i \tau )}{\beta }}-z_3\right). \nonumber
\end{align}
On implementation of the limits of integration $\tau = 0$ and $\tau = \beta$, the non logarithmic terms in the above expression vanish, while the contributions of the logarithmic terms follow the relation in \cref{log_identity}. Due to the relation in \cref{log_identity}, the limits of integration over $x$ for each term in the integrand gets modified as follows
\begin{align}
\int_{-\infty}^{\infty} dx \to \int_{\frac{\beta}{2\pi} \log z_j}^{\infty} dx, \quad \text{for}\: j=1,2,3. \nonumber
\end{align}
The integration over $x$ may now be performed to arrive at
\begin{align}
& \int_{\mathcal{M}} ~ z^2 
\left[ \frac{1}{\left(z-z_1\right)^2}+\frac{1}{\left(z-z_2\right)^2}
+\frac{1}{\left(z-z_3\right)^2}
+\frac{\left(-3 z+z_1+z_2+z_3\right) }{\left(z-z_1\right) \left(z-z_2\right) \left(z-z_3\right)} \right] \\
& = \frac{\beta ^2}{2\pi} \left[ \frac{\left(z_1^2-z_2 z_3\right) \log \left(\frac{z_1}{z_2}\right)}{z_{12} z_{13}}+\frac{\left(z_1 z_2-z_3^2\right) \log \left(\frac{z_2}{z_3}\right)}{z_{23}z_{13}}\right]. \nonumber
\end{align}
As earlier, the anti holomorphic part of the integral gives result identical to the holomorphic part.

\subsection{A single interval}
\label{app_sg}

The holomorphic part of the integral in \cref{coee_sg_int} is given by
\begin{align}
& \int_{\mathcal{M}} d^2 w  \sum_{j=1}^{4} \left(  \frac{z^2}{(z-z_j)^2}
- \frac{z^2}{(z-z_j)}\partial_{z_j}\log\left[z^{2}_{41}z^{2}_{23}\, \eta f(\eta) \right] \right) \\
& = \int_0^{\infty} dx \int_0^{\beta} d\tau \;
e^{\frac{4 \pi (x +i \tau)}{\beta}}
\left[ \sum_{j=1}^{4} \frac{1}{\left( e^{\frac{2 \pi (x +i \tau)}{\beta}}-z_j\right)^2} \right. \nonumber \\
& + \frac{-4 e^{\frac{4 \pi (x +i \tau)}{\beta}} -2z_3z_2 +z_1(-z_2+z_3-2z_4)+z_2z_4-z_3z_4+2e^{\frac{2 \pi (x +i \tau)}{\beta}}(z_1+z_2+z_3+z_4)}{\left( e^{\frac{2 \pi (x +i \tau)}{\beta}}-z_1\right)\left( e^{\frac{2 \pi (x +i \tau)}{\beta}}-z_2\right)\left( e^{\frac{2 \pi (x +i \tau)}{\beta}}-z_3\right)\left( e^{\frac{2 \pi (x +i \tau)}{\beta}}-z_4\right)}  \nonumber \\
& \left. -\frac{z_{21}z_{32}z_{41}z_{43}f'(\eta)}{\left( e^{\frac{2 \pi (x +i \tau)}{\beta}}-z_1\right)\left( e^{\frac{2 \pi (x +i \tau)}{\beta}}-z_2\right)\left( e^{\frac{2 \pi (x +i \tau)}{\beta}}-z_3\right)\left( e^{\frac{2 \pi (x +i \tau)}{\beta}}-z_4\right)z_{31}z_{42} f(\eta)} \right] . \nonumber
\end{align}
The indefinite integration over $\tau$ gives
\begin{align}
\label{tau_sg}
\frac{i \beta}{2 \pi} \sum_{j=1}^{4} \left[ B_j+ C_j \, \log \left( e^{\frac{2 \pi (x+i \tau)}{\beta}}-z_j  \right) \right] ,
\end{align}
where
\begin{align}
\label{sg_non_log}
B_j=\frac{z_j}{e^{\frac{2 \pi (x+i\tau)}{\beta}}-z_j}\;,\qquad j=1,2,3,4,
\end{align}
and $C_1, C_2, C_3 $ and $C_4$ are given as follows
\begin{align}
C_1 & = -\frac{1}{z_{31}^2} \left[ \frac{z_{31} (z_1^3+z_1^2(z_4-2z_3)+ z_1 z_2 (z_3-2 z_4)+ z_2 z_3 z_4 )}{z_{41} z_{21}} + \frac{z_1 z_{32} z_{43} f'(\eta)}{z_{42} f(\eta)}   \right] , \\
C_2 & = \frac{1}{z_{42}^2} \left[ \frac{z_{42} (z_2 ^3+ z_2 ^2 (z_3-2z_4)+z_1 z_2 (z_4-2z_3)+ z_1 z_3 z_4 )}{z_{32} z_{21}} + \frac{z_2 z_{41} z_{43} f'(\eta)}{z_{31} f(\eta)}   \right] , \nonumber \\
C_3 & = -\frac{1}{z_{31}^2 } \left[ \frac{z_{31} (z_3^3+(z_2-2 z_1) z_3^2+(z_1-2 z_2) z_4 z_3+z_1 z_2 z_4)}{z_{43}z_{32}} +\frac{z_3z_{21} z_{41} f'(\eta)}{z_{42} f(\eta)} \right] , \nonumber \\
C_4 & = \frac{1}{z_{42}^2} \left[ \frac{z_{42} (z_4^3+(z_1-2 z_2) z_4^2+(z_2-2 z_1) z_3 z_4+z_1 z_2 z_3)}{z_{41} z_{43}} +\frac{z_4z_{21} z_{32} f'(\eta)}{z_{31} f(\eta)}    \right] . \nonumber
\end{align}
Once again the non logarithmic terms described by \cref{sg_non_log} vanish on insertion of the limits of integration $\tau = 0$ and $\tau= \beta$, whereas the logarithmic terms in \cref{tau_sg} contribute according to the relation in \cref{log_identity}, which modifies the limits of the integration over $x$ as follows
\begin{align}
\int_{-\infty}^{\infty} dx \to \int_{\frac{\beta}{2\pi}\log z_j}^{\infty} dx, \quad j=1,2,3,4.
\end{align}
The integration over $x$ for the integrand in \cref{tau_sg} may now be performed with the modified limits described above to arrive at
\begin{align}
-\frac{\beta^2}{2 \pi} \sum_{j=1}^{4} C_j \log z_j \; .
\end{align}
The desired correction to the OEE of a single interval of length $\ell$ may now be obtained through the substitutions 
$\{z_1, z_2, z_3, z_4\} \to \{e^{-\frac{2\pi L}{\beta}}, e^{-\frac{2\pi \ell}{\beta}}, 1, e^{\frac{2\pi L}{\beta}}\}$
and subsequent implementation of the bipartite limit $L\to\infty$ as follows
\begin{align}
& \lim_{L\to\infty}\int_{\mathcal{M}} d^2 w \; \sum_{j=1}^{4}  \left[\frac{z^2}{(z-z_j)^2} - \frac{z^2}{(z-z_j)}\partial_{z_j}\log\left[z^{2}_{23} z_{41}^2 \eta f(\eta)\right]\right] \\
& = \ell \beta \left(-\frac{1}{\left( e^{\frac{2 \pi \ell}{\beta}} -1  \right)}
+ e^{-\frac{2 \pi \ell}{\beta}} \frac{ f' \left[ e^{-\frac{2 \pi \ell}{\beta}} \right]}
{2 f \left[ e^{-\frac{2 \pi \ell}{\beta}} \right]}  \right)
- \lim_{L\to\infty} \left[  L \beta 
\coth \left( \frac{2 \pi L}{\beta} \right)  \right]. \nonumber
\end{align}
As before the anti holomorphic part of the integral produces identical result to the holomorphic part.

\bibliographystyle{utphys}
\bibliography{reference}

\end{document}